%% file: main.tex
\documentclass[preprint,review,12pt]{elsarticle} 

\journal{ISA Transactions}

\usepackage[utf8]{inputenc}
\input{definitions.tex}
\usepackage{graphicx}
\usepackage{multicol}
\usepackage{amsmath}
\usepackage{amssymb}
\usepackage{color}
\usepackage{epstopdf}
\usepackage{bm}
\usepackage{mathtools}
\usepackage{array}
\usepackage{adjustbox}
\usepackage{lscape}
\usepackage{floatpag}

\usepackage{lineno,hyperref}
\modulolinenumbers[5]

\usepackage{diagbox}
\usepackage{multirow}
\def\rot#1{\rotatebox{90}{#1}}

\newtheorem{remark}{Remark}
\newtheorem{assumption}{Assumption}

\begin{document}

\begin{frontmatter}

\title{Cascade Extended State Observer\\ for \color{black} Active Disturbance Rejection Control \color{black}\\ Applications under Measurement Noise}

\author[PUT]{Krzysztof Łakomy\corref{mycorrespondingauthor}}
\ead{krzysztof.lakomy92@gmail.com}

\author[JNU]{Rafal Madonski\corref{mycorrespondingauthor}}
\cortext[mycorrespondingauthor]{Corresponding author}
\ead{rafal.madonski@jnu.edu.cn}

\address[PUT]{Institute of Automation and Robotics, Poznań University of Technology,\\Piotrowo 3A, 60-965, Poznań, Poland}
\address[JNU]{Energy Electricity Research Center, International Energy College, Jinan University, Zhuhai, Guangdong, 519070 P.~R.~China}


\begin{abstract}
\color{black}
The extended state observer (ESO) plays an important role in the design of feedback control for nonlinear systems. However, its high-gain nature creates a challenge in engineering practice in cases where the output measurement is corrupted by non-negligible, high-frequency noise. The presence of such noise puts a constraint on how high the observer gains can be, which forces a trade-off between fast convergence of state estimates and quality of control task realization. In this work, a new observer design is proposed to improve the estimation performance in the presence of noise. In particular, a unique cascade combination of ESOs is developed, which is capable of fast and accurate signals reconstruction, while avoiding over-amplification of the measurement noise. The effectiveness of the introduced observer structure is verified here while working as a part of an active disturbance rejection control (ADRC) scheme. The conducted numerical validation and theoretical analysis of the new observer structure show improvement over standard solution in terms of noise attenuation.
\color{black}
\end{abstract}


\end{frontmatter}


\section{Introduction}

A common strategy in the class of active disturbance rejection control (ADRC~\cite{Han-fromPID,Gao-centrality}) techniques for extracting otherwise unavailable information about the governed system is to use an extended state observer (ESO). It simultaneously provides information about the missing state variables (required for controller synthesis) as well as the lumped disturbances/uncertainties (also denoted as \textit{total disturbance}~\cite{Concept}). Successful deployments of ADRC have been well documented in various control areas including process \cite{HeTingUSaa,PowerPlantSun}, power \cite{SunliiPower,LiliDong}, and motion \cite{KLEJC2020,PATELSKI2020,ADRCnonhol,ISAT-exp1,michalek2019,BallTr} control. The effectiveness of the ADRC methodology has been also validated theoretically addressing its applicability to linear time-invariant (LTI)~\cite{Wenchao-LTI} and nonlinear time-variant (NTV)~\cite{boundedDist,Wenchao-NTV} systems by using various tools like Lyapunov~\cite{StabLyapunov} and singular perturbation techniques~\cite{Gao-stabilitySingPert}. \textcolor{black}{A recent review~\cite{BaoReview} covers standard and new theoretical achievements in ADRC for uncertain finite-dimensional and infinite-dimensional systems.} To date, different ESO architectures have been proposed and tested to address specific control problems (e.g.~\cite{KLESOarchit,RMsurvey}). The motivation for studying ADRC also comes from the fact that it is one of just few class of control approaches that was successfully transitioned from academia to industry on a significant scale and is being used in commercial motion and process control products (e.g. from Danfoss, Texas Instruments).



%

To a large extent, the performance of any ADRC scheme relies on the speed and accuracy of the ESO. It is well-known that high-gain observers (including ESO) are robust against model uncertainty and disturbances. However, the theory of observers also reveals the existence of a trade-off between speed/accuracy of state reconstruction and sensitivity to high-frequency measurement noise~\cite{khalil2014}. \textcolor{black}{The adverse, non-negligible effects of measurement noise are omnipresent in engineering practice and can be found in various important applications, including fuel cells~\cite{SunLiFuela}, magnetic levitation-based transportation~\cite{WeiMagnLev}, water management systems~\cite{positivity}, and aerial vehicles~\cite{aguilar2018}. The sensor noise is also an important factor in the design of effective adaptive algorithms including adaptive controllers~\cite{ReviewTurek1} and adaptive parameter estimators~\cite{ReviewTurek2}.} Hence, the problem of noise is still an active research topic with different solutions proposed to attenuate the effects of measurement noise to date. They mainly address the problem by: employing nonlinear \cite{Han-fromPID,KhalinNonlineaObs} or adaptive techniques \cite{KhalilSwitchin,adaptiveGainPraly,balotelli}, redesigning the local behavior by combining different observers \cite{combstrobs1,combstrobs2,LMIHG,WenchaoESFSunLi}, employing low-power structures \cite{herbst2020halfgain,ESOMarconi,limiPowerhighga,KhalilCascade},
increasing the observer order with integral terms~\cite{PIO,piointegralExp}, adding special saturation functions~\cite{Stubborn}, or modifying standard low-pass filters~\cite{OutputInjecFilt,KhallLowwPAs}.

\color{black}

The above techniques, although suitable for certain control scenarios, in general, may encounter several limitations from theoretical and/or practical point of view. Such limitations include complicated implementation (nonlinear techniques), additional assumptions of persistent excitation of certain signals (adaptive techniques), extra knowledge requirement for the design and tuning of additional components (combination of different observers), increased noise attenuation but reduced signal convergence speed (low-power structures), introduction of nonlinear components in the design which complicate its theoretical analysis and practical implementation (saturation functions), and introduction of extra system lag (modifications of standard low-pass filters).

\color{black}

Motivated by the above limitations and the practical importance of an effective denoising algorithm in nonlinear feedback control, in this work, a new paradigm to redesign high-gain observers is proposed in order to improve their performance in the presence of measurement noise. It directly addresses the limitations posed by the trade-off between speed/accuracy of state estimation and noise sensitivity. The principle behind the proposed approach is based on a decomposition of the unknown total disturbance into a predefined number of parts, each representing certain signal frequency range, and reconstruction of the decomposed parts with a set of cascaded observers. In contrast to a standard, single-observer ADRC, the introduced multi-observer topology allows to use higher observer bandwidths for reconstructing signals with smaller sensor noise impact. Hence, the contribution of this work is a proposition of the new cascade ESO topology, which is capable of providing fast and accurate estimates while avoiding over-amplification of the sensor noise and retaining the relatively straightforward implementation. To the best of our knowledge, such design is presented for the first time in the context of current literature.


\textit{Notation.} Throughout this paper, we use $\realNumbers$ as a set of real numbers, $\positiveRealNumbers=\{x\in\realNumbers:x>0\}$ as a set of positive real numbers, $\integerNumbers$ as a set of integers, while $\zeroMatrix$ and $\identityMatrix$ represent zero and identity matrices of the appropriate order, respectively. Relation $\pmb{A}\succ0$ means that $\pmb{A}$ is positive-definite, $\module{\pmb{x}}$ corresponds to the Euclidean norm of the vector $\pmb{x}$, $\module{\pmb{A}}$ is a matrix norm defined as $\module{\pmb{A}}\triangleq\sup\{\module{\pmb{Ax}}:\pmb{x}\in\realNumbers^n \ \textrm{and} \module{\pmb{x}}=1\}$, while  $\minEigenvalue{\pmb{A}}$ and $\maxEigenvalue{\pmb{A}}$ correspond to the minimal and maximal eigenvalues of matrix $\pmb{A}$. Set $\differentiableFunctionSet{1}$ represents a class of locally Lipschitz continuously differentiable functions, while $\kappaFunction$ is a class of strictly increasing functions with the zero-value at the origin.

\section{\textcolor{black}{Preliminaries: standard ESO}}
\label{sec:ConventionalESO}

Before the proposed cascade ESO is explained, theoretically proven, and finally validated in a numerical simulation, let us first recall the standard ESO and highlight its limitations when used in noisy environments. First, a nonlinear   dynamical system is represented by the state-space model
\begin{align}
    \begin{cases}
        \stateVectorDerivative(t) = \stateMatrix{n}\stateVector(t)+\inputVector{n}(\systemNonlinearDynamics(\stateVector,t)+\controlSignalField(\stateVector,t)\controlSignal(t)+\disturbance(t)), \\
        \systemOutput(t) = \outputVector{n}^\top\stateVector(t)+\measurementNoise(t),
    \end{cases}\hspace{-0.25cm}
    \label{eq:systemModel}
\end{align}
where $\stateVector \triangleq [\stateVectorElement{1} \ ... \ \stateVectorElement{n}]^\top\in\realNumbers^n$ is a system state, $\disturbance\in\realNumbers$ is an external disturbance, $\controlSignal\in\realNumbers$ is a control signal, $\systemOutput\in\realNumbers$ is a system output, $\measurementNoise\in\realNumbers$ corresponds to the measurement noise, $\controlSignalField\in\realNumbers\color{black}\textrm{\textbackslash}\{0\}\color{black}$ describes the influence of the control signal on the system dynamics, $\systemNonlinearDynamics\in\realNumbers$ represents lumped dynamics of the controlled plant, \color{black} time $t\in\positiveRealNumbers\cup\{0\}$, \color{black} while \color{black}$n$ is a known order of the dynamical system\color{black},
%
$\stateMatrix{n} \triangleq \begin{bmatrix} \zeroMatrix^{n-1 \times 1} & \identityMatrix^{n-1\times n-1} \\
0 & \zeroMatrix^{1\times n-1}\end{bmatrix}\in\realNumbers^{n\times n}$, $\inputVector{n} \triangleq \left[\zeroMatrix^{\color{black}1\times n-1\color{black}} \ 1\right]^\top\in\realNumbers^n$, and $\outputVector{n} \triangleq \left[1 \ \zeroMatrix^{\color{black}1\times  n-1\color{black}}\right]^\top\in\realNumbers^n$.

\begin{assumption}
    \label{ass:1}
    System \eqref{eq:systemModel} is defined on an arbitrarily large bounded domain
    \color{black} $\systemStateDomain\triangleq\{\stateVector\in\realNumbers^{n}:\module{\stateVector}<r_x\}$ for $r_x>0$, \color{black}
    such that $\stateVector\in\systemStateDomain$.
\end{assumption}

\begin{assumption}
    \label{ass:2}
    Measurement noise is bounded in the sense that there exists a bounded set \color{black} $\measurementNoiseDomain\triangleq\{\measurementNoise\in\realNumbers:\abs{\measurementNoise}<\measurementNoiseUpperBound\}$ for some $\measurementNoiseUpperBound>0$, \color{black} such that $\measurementNoise\in\measurementNoiseDomain$.
\end{assumption}

\begin{assumption}
    \label{ass:3}
    External disturbance $\disturbance(t)\in\differentiableFunctionSet{1}$, and 
    $\disturbance\in\externalDisturbanceDomain$, $\disturbanceDerivative\in\externalDisturbanceDerivativeDomain$ for some bounded sets \color{black} $\externalDisturbanceDomain\triangleq\{\disturbance\in\realNumbers:\abs{\disturbance}<r_{d^*}\}$ and $\externalDisturbanceDerivativeDomain\triangleq\{\disturbanceDerivative\in\realNumbers:\abs{\disturbanceDerivative}<r_{\dot{d}^*}\}$, for some $r_{d^*},r_{\dot{d}^*}>0$. \color{black}
\end{assumption}

\begin{assumption}
    \label{ass:4}
    Fields $\controlSignalField(\stateVector,t):\systemStateDomain\times\realNumbers\rightarrow\realNumbers\color{black}\textrm{\textbackslash}\{0\}\color{black}$ and $\systemNonlinearDynamics(\stateVector,t):\systemStateDomain\times\realNumbers\rightarrow\realNumbers$ are continously differentiable locally Lipschitz functions, i.e., $\controlSignalField,\systemNonlinearDynamics\in\differentiableFunctionSet{1}$.
\end{assumption}

\begin{assumption}
    \label{ass:5}
    Utilized control input $\controlSignal\in\differentiableFunctionSet{1}$, and $\sup_{t\geq0}\{|u(t)|,|\dot{u}(t)|\}<r_u$ for some $r_u>0$.
\end{assumption}

State dynamics, taken from \eqref{eq:systemModel}, can be rewritten as
\begin{align}
    \begin{cases}
        \stateVectorDerivative(t) = \stateMatrix{n}\stateVector(t)+\inputVector{n}\controlSignalFieldEstimate\controlSignal(t), \\
        \hspace{0.8cm}+ \underbrace{ \inputVector{n}(\systemNonlinearDynamics(\stateVector,t)+\disturbance(t)+(\controlSignalField(\stateVector,t)-\controlSignalFieldEstimate)\controlSignal(t))}_{\inputVector{n}\totalDisturbance(\stateVector,t)}, \\
        \systemOutput(t) = \outputVector{n}^\top\stateVector(t)+\measurementNoise(t),
    \end{cases}
\label{eq:rewrittenSystemDynamics}
\end{align}
where $\totalDisturbance(\stateVector,t):\systemStateDomain\times\realNumbers\rightarrow\realNumbers$ is the so-called total disturbance, and $\controlSignalFieldEstimate\in\realNumbers\color{black}\textrm{\textbackslash}\{0\}\color{black}$ is a rough, constant estimate of $\controlSignalField$.
%
\color{black}
\begin{remark}
    \label{rem:strojenieb}
     If the mathematical model of system~\eqref{eq:systemModel} is known, the system input gain $\hat{g}$ can be straightforwardly calculated. If not, an approximated value of $\hat{g}$ can still suffice as ADRC is robust (to some extent) against parametric uncertainty. The necessary and sufficient condition for the uncertain design parameter $\hat{g}$ has been established in~\cite{chen2020necessary}.
\end{remark}
\color{black}

Now, let us define the extended state $\extendedState\triangleq[\stateVector^\top \ \totalDisturbance]^\top\in\realNumbers^{n+1}$ with the dynamics expressed, according to \eqref{eq:rewrittenSystemDynamics}, as
\begin{align}
    \begin{cases}
        \extendedStateDerivative(t) = \stateMatrix{n+1}\extendedState(t) +\inputVector{n+1}\totalDisturbanceDerivative(\extendedState,t) + \inputVectorExtendedState{n+1}\controlSignalFieldEstimate\controlSignal(t), \\ \systemOutput(t) = \outputVector{n+1}^\top\extendedState(t) + \measurementNoise(t),
    \end{cases}
    \label{eq:extendedStateSystem}
\end{align}
%
 where $\inputVectorExtendedState{n+1} \triangleq \left[\zeroMatrix^{\color{black}1\times n-1\color{black}} \ 1 \ 0\right]^\top\in\realNumbers^{n+1}$.

The standard ESO design, see e.g. \cite{Gao-pole-placement}, is expressed by the dynamics of the extended state estimate $\extendedStateEstimate:=\observerStageState{1}\in\realNumbers^{n+1}$, i.e.,
%
\begin{align}
    \observerStageStateDerivative{1} &= \stateMatrix{n+1}\observerStageState{1}+\inputVectorExtendedState{n+1}\controlSignalFieldEstimate\controlSignal+\observerGainVector{1,n+1}(\systemOutput-\outputVector{n+1}^\top\observerStageState{1}),
    \label{eq:cascadedObserverFirstStageDynamics}
\end{align}
where $\observerGainVector{1,n+1}\triangleq[\observerGainVectorElementCoefficient{1}\observerStageBandwidth{1} \ ... \ \observerGainVectorElementCoefficient{n+1}\observerStageBandwidth{1}^{n+1}]^\top\in\realNumbers^{n+1}$ is the observer gain vector dependent on the coefficients $\observerGainVectorElementCoefficient{i}\in\positiveRealNumbers$ for ${i\in\{1,...,n+1\}}$ and on the parameter $\observerStageBandwidth{1}\in\positiveRealNumbers$. Observation error can be defined as $\observerStageObservationError{1}\triangleq\extendedState-\observerStageState{1}\in\realNumbers^{n+1}$, while its dynamics, derived upon \eqref{eq:extendedStateSystem} and \eqref{eq:cascadedObserverFirstStageDynamics}, can be expressed as
%
%
\begin{align}
    \observerStageObservationErrorDerivative{1} = \underbrace{(\stateMatrix{n+1}-\observerGainVector{1,n+1}\outputVector{n+1})}_{\observationErrorStateMatrix{1}}\color{black}\observerStageObservationError{1}\color{black}+\inputVector{n+1}\totalDisturbanceDerivative-\observerGainVector{1,n+1}\measurementNoise.
    \label{eq:firstStageObservationError}
\end{align}
\begin{remark}
    \label{remPP}
    For the sake of notation conciseness of further analysis, we propose to choose the values of $\observerGainVectorElementCoefficient{i}$ in a way to obtain matrix $\observationErrorStateMatrix{1}$ with all eigenvalues equal to $-\observerStageBandwidth{1}$.
\end{remark}
\begin{remark}
        \label{rem:2}
        \color{black} Assumptions \ref{ass:1}-\ref{ass:5} imply bounded values of the total disturbance $\totalDisturbance\in\differentiableFunctionSet{1}$, and its derivative  $\totalDisturbanceDerivative\in\totalDisturbanceDerivativeDomain$ \color{black} for $\totalDisturbanceDerivativeDomain\triangleq\{\totalDisturbanceDerivative\in\realNumbers:\abs{\totalDisturbance}<\totalDisturbanceDerivativeUpperBound\}$, where $\totalDisturbanceDerivativeUpperBound>0$\color{black}.
        The need to ensure bounded values of $\totalDisturbanceDerivative$, interpreted as the perturbation of the observation error system \eqref{eq:firstStageObservationError}, is a well-known characteristics of the disturbance-observer-based controllers, recently discussed in depth in~\cite{Concept}. This limitation has been extensively discussed in the area of the ADRC techniques to date. Latest theoretical developments in the field have shown under what conditions (which turn out not to be difficult to satisfy in practice) it can be proved that the total disturbance $d(\bm{x},t)$ and its consecutive time-derivatives will remain bounded even if it is a function of system states. This property is satisfied with the control input having a sufficiently large upper bound (see \cite{Zhaoproof}), what was imposed by Assumption~\ref{ass:5}. Additionally, when a system state is within some bounded set, imposed by Assumption~\ref{ass:1}, the boundedness of state-dependent disturbance $\totalDisturbance(\stateVector,t)$ is satisfied (see \cite{SemiGlobalJun}).
    \color{black}
\end{remark}

%




Now, in order to show the influence of total disturbance and measurement noise on the observation errors for standard ESO, let us first introduce a linear change of coordinates   $\observerStageObservationError{1}\triangleq\stateTransformationMatrix{1}\transformedState{1}$, where $\stateTransformationMatrix{1}\triangleq\textrm{diag}\{\observerStageBandwidth{1}^{-n},...,\observerStageBandwidth{1}^{-1},1\}\in\realNumbers^{n+1\times n+1}$ and $\transformedState{1}\in\realNumbers^{n+1}$, resulting in the transformation of~\eqref{eq:firstStageObservationError} to
\begin{align}
    \transformedStateDerivative{1} &= \stateTransformationMatrix{1}^{-1}\observationErrorStateMatrix{1}\stateTransformationMatrix{1}\transformedState{1} + \stateTransformationMatrix{1}^{-1}\inputVector{n+1}\totalDisturbanceDerivative-\stateTransformationMatrix{1}^{-1}\observerGainVector{1,n+1}\measurementNoise \nonumber \\
    &= {\observerStageBandwidth{1}\transformedStateStateMatrix{1}}\transformedState{1} + \inputVector{n+1}\totalDisturbanceDerivative-\stateTransformationMatrix{1}^{-1}\observerGainVector{1,n+1}\measurementNoise,
    \label{eq:transformedStateDynamicsFirstStage}
\end{align}
where
\begin{align}
    \transformedStateStateMatrix{i} = \begin{bmatrix}-\observerGainVectorElementCoefficient{1} & 1  & \cdots & 0 \\
         \vdots & \vdots & \ddots & \vdots \\
        -\observerGainVectorElementCoefficient{n}  & 0 & \cdots & 1 \\
        -\observerGainVectorElementCoefficient{n+1}  & 0 & \cdots & 0\end{bmatrix}.
        \label{eq:hStarMatrix}
\end{align}
We propose a Lyapunov function candidate $\observerStageLyapunovFunction{1} = \transformedState{1}^\top\lyapunovEquationSolutionMatrix{1}\transformedState{1}$, $\observerStageLyapunovFunction{1}:\realNumbers^{n+1}\color{black}\rightarrow\realNumbers$ limited by $\minEigenvalue{\lyapunovEquationSolutionMatrix{1}}\module{\transformedState{1}}\leq\observerStageLyapunovFunction{1}\leq\maxEigenvalue{\lyapunovEquationSolutionMatrix{1}}\module{\transformedState{1}}$,
where $\lyapunovEquationSolutionMatrix{1}\succ0$ is the solution of Lyapunov equation
\begin{align}
{\transformedStateStateMatrix{1}}^\top\lyapunovEquationSolutionMatrix{1}+\lyapunovEquationSolutionMatrix{1}\transformedStateStateMatrix{1} = -\identityMatrix.
\label{eq:lyapunovEquation}
\end{align}
The derivative of $\observerStageLyapunovFunction{1}$, derived upon \eqref{eq:transformedStateDynamicsFirstStage}, can be expressed as
\begin{align}
    \observerStageLyapunovFunctionDerivative{1} &= -\observerStageBandwidth{1}\transformedState{1}^\top\transformedState{1} + 2\transformedState{1}^\top\lyapunovEquationSolutionMatrix{1}\inputVector{n+1}\totalDisturbanceDerivative-2\transformedState{1}^\top\lyapunovEquationSolutionMatrix{1}\stateTransformationMatrix{1}^{-1}\observerGainVector{1,n+1}\measurementNoise \nonumber \\
    &\leq - \observerStageBandwidth{1}\module{\transformedState{1}}^2+2\module{\lyapunovEquationSolutionMatrix{1}}\module{\transformedState{1}}\abs{\totalDisturbanceDerivative}+2\module{\lyapunovEquationSolutionMatrix{1}}\maximalGainCoefficient\observerStageBandwidth{1}^{n+1}\color{black}\abs{\measurementNoise},\color{black}
\end{align}
for $\maximalGainCoefficient=\max_j({\observerGainVectorElementCoefficient{j}})$.
The derivative of $\observerStageLyapunovFunction{1}$ holds
\begin{align}
    \observerStageLyapunovFunctionDerivative{1} &\leq  - \observerStageBandwidth{1}(1-\observerStageMajorizationConstant{1})\module{\transformedState{1}}^2 \textrm{for} \nonumber \\
    \module{\transformedState{1}} &\geq \frac{2\module{\lyapunovEquationSolutionMatrix{1}}}{\observerStageBandwidth{1}\observerStageMajorizationConstant{1}}\abs{\totalDisturbanceDerivative}+\frac{2\module{\lyapunovEquationSolutionMatrix{1}}\maximalGainCoefficient\observerStageBandwidth{1}^{n}}{\observerStageMajorizationConstant{1}}\abs{\measurementNoise},
    \label{eq:firstStageLyapunovInequality}
\end{align}
where $\observerStageMajorizationConstant{1}\in(0,1)$ is a chosen majorization constant and the conservatively estimated lower bound of $\module{\transformedState{1}}$ is a class $\kappaFunction$ function with respect to arguments $\abs{\totalDisturbanceDerivative}$ and $\abs{\measurementNoise}$. According to the result \eqref{eq:firstStageLyapunovInequality}, Assumption \ref{ass:2}, and Remark \ref{rem:2}, we can claim that system \eqref{eq:transformedStateDynamicsFirstStage} is input-to-state stable\footnote{\color{black}The definition and selected properties of input-to-state stable system are described in the Appendix A.\color{black}}
(ISS) with respect to perturbations $\abs{\totalDisturbanceDerivative}$ and $\abs{\measurementNoise}$, and satisfies
\begin{align}
    \limsup_{t\rightarrow\infty}\module{\transformedState{1}(t)} &\leq \issFactor\frac{2\module{\lyapunovEquationSolutionMatrix{1}}}{\observerStageBandwidth{1}\observerStageMajorizationConstant{1}}\sup_{t\geq0}\abs{\totalDisturbanceDerivative(t)} \nonumber \\
    &+\issFactor\frac{2\module{\lyapunovEquationSolutionMatrix{1}}\maximalGainCoefficient\observerStageBandwidth{1}^{n}}{\observerStageMajorizationConstant{1}}\sup_{t\geq0}\abs{\measurementNoise(t)} \nonumber \\
    &\leq \issFactor\frac{2\module{\lyapunovEquationSolutionMatrix{1}}}{\observerStageBandwidth{1}\observerStageMajorizationConstant{1}}\totalDisturbanceDerivativeUpperBound+\issFactor\frac{2\module{\lyapunovEquationSolutionMatrix{1}}\maximalGainCoefficient\observerStageBandwidth{1}^{n}}{\observerStageMajorizationConstant{1}}\measurementNoiseUpperBound,
    \label{eq:firstStageCascadeResult}
\end{align}
for $\issFactor = \sqrt{\maxEigenvalue{\lyapunovEquationSolutionMatrix{1}}/\minEigenvalue{\lyapunovEquationSolutionMatrix{1}}}$.
To achieve the minimal asymptotic upper-bound of the transformed observation error $\module{\transformedState{1}}$, we need to find a trade-off between the reduction of $\abs{\totalDisturbanceDerivative}$ impact with the increasing $\observerStageBandwidth{1}$ values and the influence of $\abs{\measurementNoise}$ amplified by $\observerStageBandwidth{1}^{n}$. It is worth noting, that for the nominal case when $\totalDisturbanceDerivative(t)\equiv0$ and $\measurementNoise(t)\equiv0$, the asymptotic upper bound of $\limsup_{t\rightarrow\infty}\module{\transformedState{1}}=0$, implying the asymptotic upper bound of the original observation error vector $\limsup_{t\rightarrow\infty}\module{\observerStageObservationError{1}}=0$. In the case of non-zero values of $\abs{\totalDisturbanceDerivative}$ and $\measurementNoise(t)\equiv0$, $\module{\transformedState{1}}\rightarrow0$ as $\observerStageBandwidth{1}\rightarrow\infty$ and $t\rightarrow\infty$.

\begin{remark}
     \label{rem:3}
     Since in the practical conditions the value of parameter $\observerStageBandwidth{1}\gg1$, inequality $\module{\observerStageObservationError{1}}\leq\maxEigenvalue{\stateTransformationMatrix{1}}\module{\transformedState{1}}$ implies that the asymptotic relation \eqref{eq:firstStageCascadeResult} and all of the further comments hold also for the original observation error $\module{\observerStageObservationError{1}}$.
\end{remark}



\section{\textcolor{black}{Main result: cascade ESO}}
\label{sec:concept}

\subsection{\textcolor{black}{Concept}}

\color{black} Following~\cite{Gao-centrality}, \color{black} the concept of ADRC relies on the feedforward cancellation of the total disturbance included in the generic controller
 \begin{align}
     \controlSignal = \controlSignalFieldEstimate^{-1}(-\totalDisturbanceEstimate + \virtualControlSignal{1}),
     \label{eq:controlSignalSimple}
 \end{align}
 where $\virtualControlSignal{1}$ represents the new virtual control signal, most commonly in the form of a stabilizing feedback controller. \color{black} After the application of control signal $\controlSignal$ into dynamics \eqref{eq:rewrittenSystemDynamics}, one can observe that the first component of \eqref{eq:controlSignalSimple}, i.e. $-\controlSignalFieldEstimate\totalDisturbanceEstimate$, should cancel out the total disturbance $\totalDisturbance$, while the second component, i.e. $\controlSignalFieldEstimate\virtualControlSignal{1}$,  is responsible for pushing the state vector $\stateVector$ towards the desired reference value. \color{black}

 \color{black}
 \begin{remark}
      To keep the article focused on the design of cascade ESO, we do not introduce nor analyze here any specific feedback controller $\virtualControlSignal{1}$. At this point, we assume that $\virtualControlSignal{1}:=\virtualControlSignal{}(\stateVector,\cdot)\in\differentiableFunctionSet{1}$ is some bounded and locally Lipschitz function consistent with Assumption \ref{ass:5}, that assures the asymptotic convergence of system state $\stateVector$ to its reference value for the ideal case, when $\totalDisturbanceEstimate\equiv\totalDisturbance$ \color{black} and $\measurementNoise\equiv0$.   \color{black}
 \end{remark}
 \color{black}

 Substitution of \eqref{eq:controlSignalSimple} into the dynamics \eqref{eq:rewrittenSystemDynamics} results in the closed-loop form of the considered system expressed as
\begin{align}
    \stateVectorDerivative = \stateMatrix{n}\stateVector+\inputVector{n}\totalDisturbanceObservationError+\inputVector{n}\virtualControlSignal{1},
\end{align}
where $\totalDisturbanceObservationError = \totalDisturbance - \totalDisturbanceEstimate\in\realNumbers$ is the residual total disturbance. Let us now assume that the parameter $\observerStageBandwidth{1}$ of the standard ESO, see \eqref{eq:cascadedObserverFirstStageDynamics}, was set to a relatively low value which only allows a precise following of the first element of extended state, i.e. $\outputVector{n+1}^\top\extendedState$, but filters out the measurement noise $\measurementNoise(t)$.
Latter elements of the extended state are dependent on the further derivatives of the first one, and usually are more dynamic and have faster transients, thus are not estimated precisely by ESO using chosen $\observerStageBandwidth{1}$. As a consequence, values of the total disturbance residue $\totalDisturbanceObservationError$ can be substantial causing a possible loss of control precision.
%

\begin{figure}[hbt!]
 \centering
 \includegraphics[width=0.75\textwidth]{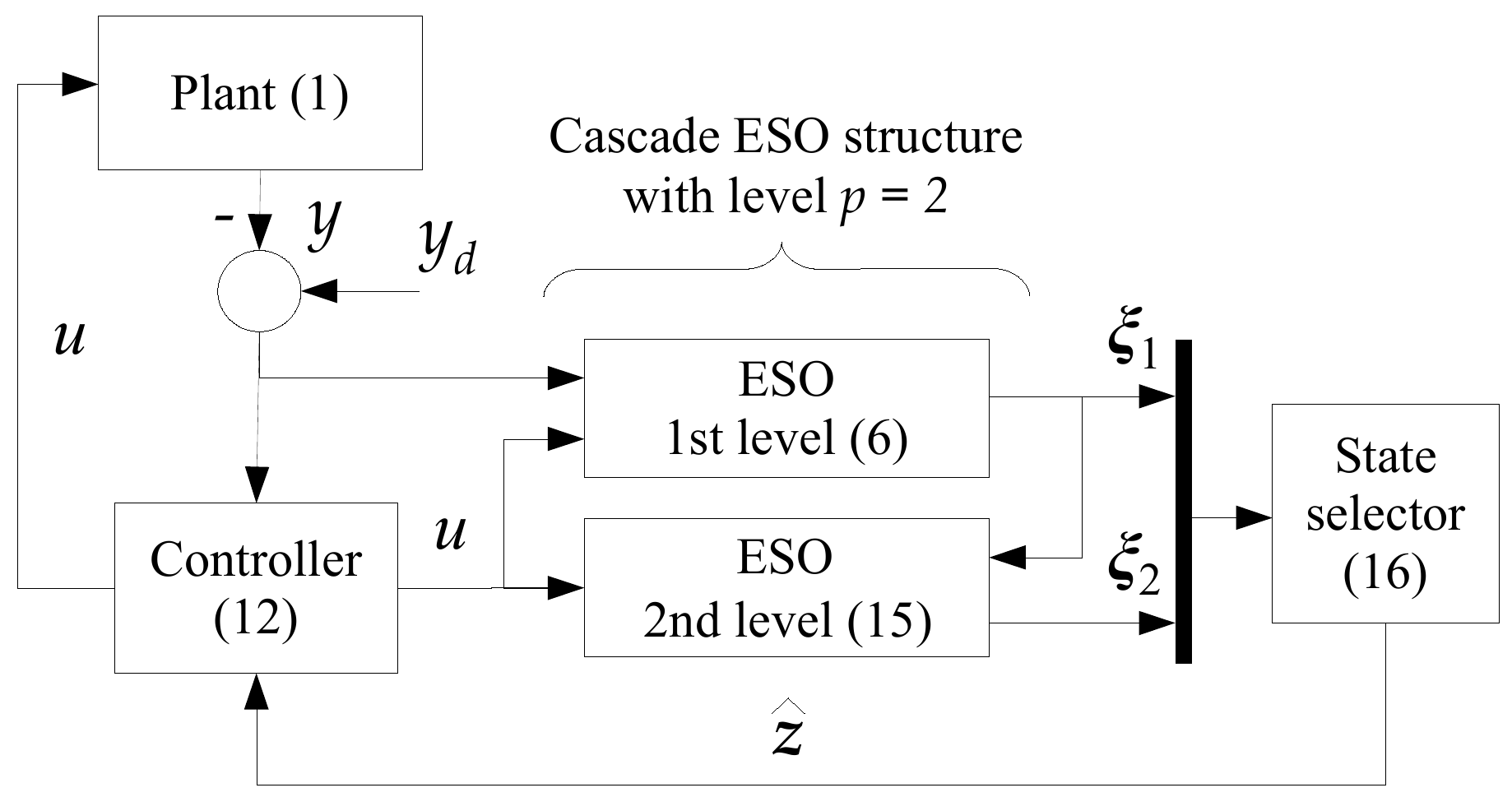}
 \caption{{Block diagram of an ADRC structure including proposed cascade ESO for $p=2$.}}
 \label{fig:BlockDiagramISATspecificCascade}
\end{figure}

To improve the estimation performance of the extended state obtained with the standard ESO with low $\observerStageBandwidth{1}$, treated here as a first level in the cascade observer structure, let us introduce the state vector of the second level of the observer $\observerStageState{2}\triangleq[\observerStageStateElementEstimate{1,1} \ \widehat{\observerStageStateElement{1,1}^{(1)}} \  ... \ \widehat{{\observerStageStateElement{1,1}^{(n-1)}}} \ \totalDisturbanceObservationErrorEstimate]^\top\in\realNumbers^{n+1}$, where $\widehat{{\observerStageStateElement{1,1}^{(i)}}}$ for $i\in\{1,..,n-1\}$ is the estimate of $i$-th derivative of $\observerStageStateElement{1,1}$, while $\totalDisturbanceObservationErrorEstimate$ is the estimated value of residual total disturbance. The structure of the second observer level is
\color{black}
\begin{equation}
    \observerStageStateDerivative{2}(t) = \stateMatrix{n+1}\observerStageState{2}(t) + \inputVectorExtendedState{n+1}\virtualControlSignal{1}(t)+\observerGainVector{2,n+1}\outputVector{n+1}^\top(\observerStageState{1}(t)-\observerStageState{2}(t)),
\end{equation}
where $\observerGainVector{2,n+1}\triangleq[\observerGainVectorElementCoefficient{1}\observerStageBandwidth{2} \ ... \ \observerGainVectorElementCoefficient{n+1}\observerStageBandwidth{2}^{n+1}]^\top\in\realNumbers^{n+1}$ and $\observerStageBandwidth{2}=\observerBandwidthMultiplier\observerStageBandwidth{1}$, $\observerBandwidthMultiplier>1$. Furthermore, by substituting the virtual control signal $\virtualControlSignal{1}(t)$ in the above equation with $\virtualControlSignal{1}(t) = \controlSignalFieldEstimate\controlSignal(t)+\totalDisturbanceEstimate(t)$ (calculated upon~\eqref{eq:controlSignalSimple}), one obtains
\begin{align}
    \observerStageStateDerivative{2}(t) &= \stateMatrix{n+1}\observerStageState{2}(t) + \inputVectorExtendedState{n+1}\overbrace{(\controlSignalFieldEstimate\controlSignal(t)+\inputVector{n+1}^\top\observerStageState{1}(t))}^{v(t)} \nonumber \\
    &+\observerGainVector{2,n+1}\outputVector{n+1}^\top(\observerStageState{1}(t)-\observerStageState{2}(t)),
    \label{eq:cascadedObserverSecondStageDynamics}
\end{align}
where $\inputVector{n+1}^\top\observerStageState{1}(t)=:\totalDisturbanceEstimate(t)$ results from the estimates of total disturbance obtained with the first cascade level \eqref{eq:cascadedObserverFirstStageDynamics}.
\color{black}
Note that the equation of the second observer level does not depend on the system output $\systemOutput$, thus is not affected by the measurement noise directly. The estimate of total disturbance utilized in \eqref{eq:controlSignalSimple} should be now taken from the new extended state estimate
\begin{align}
\extendedStateEstimate := \observerStageState{2} + \inputVector{n+1}\inputVector{n+1}^\top\observerStageState{1}.
\end{align}
%
{A block diagram of the ADRC control structure with the proposed cascade ESO (for $p=2$) applied to the system \eqref{eq:systemModel} is shown in Fig.~\ref{fig:BlockDiagramISATspecificCascade}.}

%
\begin{remark}
 Coefficients $\observerGainVectorElementCoefficient{i}$ of the cascade observer can differ in general between the cascade levels, but for the sake of notation clarity, we assume within this article that they are equal.
\end{remark}
\begin{remark}
     The idea of the cascade observer topology is to estimate the total disturbance observation residue resulting from the previous level with the next cascade level of ESO, implying that the observer bandwidth multiplier $\observerBandwidthMultiplier>1$
\end{remark}

The observation error, associated with the cascade observer with two levels, can be defined as
%
\begin{align}
    \observerStageObservationError{2} = \extendedState - \observerStageState{2} - \inputVector{n+1}\inputVector{n+1}^\top\observerStageState{1},
\end{align}
with its dynamics derived upon \eqref{eq:cascadedObserverSecondStageDynamics}, \eqref{eq:cascadedObserverFirstStageDynamics}, and \eqref{eq:extendedStateSystem} as
\begin{align}
    \observerStageObservationErrorDerivative{2}
    &= \observationErrorStateMatrix{2}\observerStageObservationError{2} + \residualObservationErrorMatrix{2,1}\observerStageObservationError{1} + \measurementNoiseVector{2}\measurementNoise + \inputVector{n+1}\totalDisturbanceDerivative,
\end{align}
for $\observationErrorStateMatrix{2} = \stateMatrix{n+1}-\observerGainVector{2,n+1}\outputVector{n+1}^\top$, $\residualObservationErrorMatrix{2,1} = \observerGainVector{2,n+1}\outputVector{n+1}^\top-\inputVector{n+1}\inputVector{n+1}^\top\observerGainVector{1,n+1}\outputVector{n+1}^\top$ and $\measurementNoiseVector{2} = -\inputVector{n+1}\inputVector{n+1}^\top\observerGainVector{1,n+1}$.

Following the reasoning presented in Sect.~\ref{sec:ConventionalESO} for standard ESO, we now prove that the observation error $\observerStageObservationError{2}$ is bounded, despite a perturbing impact of total disturbance and measurement noise. Let us introduce a linear change of coordinates $\observerStageObservationError{2}\triangleq\stateTransformationMatrix{2}\transformedState{2}$ for $\stateTransformationMatrix{2}\triangleq\textrm{diag}\{\observerStageBandwidth{2}^{-n},...,\observerStageBandwidth{2}^{-1},1\}\in\realNumbers^{n+1\times n+1}$, and $\transformedState{2}\in\realNumbers^{n+1}$. The dynamics of $\transformedState{2}$ is described by
\begin{align}
    \transformedStateDerivative{2} &= \stateTransformationMatrix{2}^{-1}\observationErrorStateMatrix{2}\stateTransformationMatrix{2}\transformedState{2} + \stateTransformationMatrix{2}^{-1}\residualObservationErrorMatrix{2,1}\stateTransformationMatrix{1}\transformedState{1}+\stateTransformationMatrix{2}^{-1}\measurementNoiseVector{2}\measurementNoise +\stateTransformationMatrix{2}^{-1}\inputVector{n+1}\totalDisturbanceDerivative \nonumber \\
    &= {\observerStageBandwidth{2}\transformedStateStateMatrix{2}}\transformedState{2} + \stateTransformationMatrix{2}^{-1}\residualObservationErrorMatrix{2,1}\stateTransformationMatrix{1}\transformedState{1}+\stateTransformationMatrix{2}^{-1}\measurementNoiseVector{2}\measurementNoise +\inputVector{n+1}\totalDisturbanceDerivative,
    \label{eq:transformedStateDynamicsSecondStage}
\end{align}
for matrix $\transformedStateStateMatrix{2}$ taken from \eqref{eq:hStarMatrix}.
Let us introduce a Lyapunov function candidate in the form $\observerStageLyapunovFunction{2} \triangleq \transformedState{2}^\top\lyapunovEquationSolutionMatrix{2}\transformedState{2}$, $\observerStageLyapunovFunction{2}:\realNumbers^{n+1} \color{black}\rightarrow\realNumbers$ limited by $\minEigenvalue{\lyapunovEquationSolutionMatrix{2}}\module{\transformedState{2}}\leq\observerStageLyapunovFunction{2}\leq\maxEigenvalue{\lyapunovEquationSolutionMatrix{2}}\module{\transformedState{2}}$,
where $\lyapunovEquationSolutionMatrix{2}$ is a solution of the Lyapunov equation \eqref{eq:lyapunovEquation}.
The derivative of $\observerStageLyapunovFunction{2}$, calculated with \eqref{eq:transformedStateDynamicsSecondStage}, has the form
\begin{align}
    \observerStageLyapunovFunctionDerivative{2} &= -\observerStageBandwidth{2}\transformedState{2}^\top\transformedState{2} + 2\transformedState{2}^\top\lyapunovEquationSolutionMatrix{2}\stateTransformationMatrix{2}^{-1}\residualObservationErrorMatrix{2,1}\stateTransformationMatrix{1}\transformedState{1}+2\transformedState{2}^\top\lyapunovEquationSolutionMatrix{2}\stateTransformationMatrix{2}^{-1}\measurementNoiseVector{2}\measurementNoise \nonumber \\
    &+ 2\transformedState{2}^\top\lyapunovEquationSolutionMatrix{2}\inputVector{n+1}\totalDisturbanceDerivative \nonumber \\
    &\leq -\observerStageBandwidth{2}\module{\transformedState{2}}^2 + 2\maximalGainCoefficient\observerBandwidthMultiplier^{n+1}\observerStageBandwidth{1}\module{\lyapunovEquationSolutionMatrix{2}}\module{\transformedState{1}}\module{\transformedState{2}}  \nonumber \\
    &+ 2\maximalGainCoefficient\observerStageBandwidth{1}^{n+1}\module{\lyapunovEquationSolutionMatrix{2}}\abs{\measurementNoise}\module{\transformedState{2}}+2\module{\lyapunovEquationSolutionMatrix{2}}\abs{\totalDisturbanceDerivative}\module{\transformedState{2}},
\end{align}
which holds
\begin{align}
    \observerStageLyapunovFunctionDerivative{2} &\leq -\observerStageBandwidth{2}(1-\observerStageMajorizationConstant{2})\module{\transformedState{2}}^2 \ \textrm{for} \nonumber \\
    \module{\transformedState{2}} &\geq \frac{2\maximalGainCoefficient\observerBandwidthMultiplier^{n}\module{\lyapunovEquationSolutionMatrix{2}}}{\observerStageMajorizationConstant{2}}\module{\transformedState{1}}+\frac{2\maximalGainCoefficient\observerStageBandwidth{1}^{n}\module{\lyapunovEquationSolutionMatrix{2}}}{\observerBandwidthMultiplier\observerStageMajorizationConstant{2}}\abs{\measurementNoise} + \frac{2\module{\lyapunovEquationSolutionMatrix{2}}}{\observerStageBandwidth{2}\observerStageMajorizationConstant{2}}\abs{\totalDisturbanceDerivative},
    \label{eq:secondStageLyapunovInequality}
\end{align}
where $\observerStageMajorizationConstant{2}\in(0,1)$ is a chosen majorization constant and the conservatively estimated lower bound of the transformed observation error $\module{\transformedState{2}}$ is a class $\kappaFunction$ function with respect to the arguments $\module{\transformedState{1}}, \ \abs{\measurementNoise},$ and $\abs{\totalDisturbanceDerivative}$. According to the result \eqref{eq:secondStageLyapunovInequality} \color{black} and the definition of ISS property described in Appendix A\color{black}, system \eqref{eq:transformedStateDynamicsSecondStage} satisfies the asymptotic relation
\begin{align}
    \limsup_{\color{black}t\rightarrow\infty\color{black}} \module{\transformedState{2}(t)} &\leq \issFactor\frac{2\maximalGainCoefficient\observerBandwidthMultiplier^{n}\module{\lyapunovEquationSolutionMatrix{2}}}{\observerStageMajorizationConstant{2}}\limsup_{t\rightarrow\infty}\module{\transformedState{1}(t)} \nonumber \\
    &+\issFactor\frac{2\maximalGainCoefficient\color{black}\observerStageBandwidth{1}^{n}\module{\lyapunovEquationSolutionMatrix{2}}}{\observerBandwidthMultiplier\observerStageMajorizationConstant{2}}\sup_{t\geq0}\abs{\measurementNoise(t)} + \frac{2\module{\lyapunovEquationSolutionMatrix{2}}}{\observerStageBandwidth{2}\observerStageMajorizationConstant{2}}\sup_{t\geq0}\abs{\totalDisturbanceDerivative(t)} \nonumber \\
    &\overset{\eqref{eq:firstStageCascadeResult}}{\leq} \issFactor\Bigg[\frac{2\maximalGainCoefficient\color{black}\observerStageBandwidth{1}^{n}\module{\lyapunovEquationSolutionMatrix{2}}}{\observerBandwidthMultiplier\observerStageMajorizationConstant{2}}+ \frac{4\maximalGainCoefficient^2\observerBandwidthMultiplier^{n}\observerStageBandwidth{1}^{n}\module{\lyapunovEquationSolutionMatrix{2}}^2}{\observerStageMajorizationConstant{2}\observerStageMajorizationConstant{1}}\Bigg]\measurementNoiseUpperBound \nonumber \\
    &+\issFactor\left[\frac{2\module{\lyapunovEquationSolutionMatrix{2}}}{\observerStageBandwidth{2}\observerStageMajorizationConstant{2}} + \frac{4\maximalGainCoefficient\observerBandwidthMultiplier^{n}\module{\lyapunovEquationSolutionMatrix{2}}^2}{\observerStageMajorizationConstant{2}\observerStageMajorizationConstant{1}\observerStageBandwidth{1}}\right]\totalDisturbanceDerivativeUpperBound,
    \label{eq:resultSecondCascadeStage}
\end{align}
for $\issFactor, \ \totalDisturbanceDerivativeUpperBound, \ \measurementNoiseUpperBound$ having the same values as the ones from \eqref{eq:firstStageCascadeResult}. Summarizing, the transformed observation error is bounded in general, $\module{\transformedState{2}}\rightarrow0$ as $\observerStageBandwidth{1}\rightarrow\infty$ and $t\rightarrow\infty$ when $\measurementNoise\equiv0$, and $\limsup_{t\rightarrow\infty}\module{\transformedState{2}(t)}=0$ when both $\measurementNoise\equiv0$ and $\totalDisturbanceDerivative\equiv0$.
Similarly to the issue described in Remark~\ref{rem:3} for the single-stage observer, the original observation error for the second-level cascade observer ($p=2$) satisfies $\module{\observerStageObservationError{2}}\leq\maxEigenvalue{\stateTransformationMatrix{2}}\module{\transformedState{2}}=\max\left\{\frac{1}{\observerStageBandwidth{2}^{-n}},1\right\}\module{\transformedState{2}}$, and thus holds the ISS property itself. \color{black} The obtained result presents practical stability achieved for the non-zero measurement noise and total disturbance derivative which are the perturbations of the analyzed system.\color{black}

\color{black}

\subsection{\textcolor{black}{General design}}
\label{sec:GeneralizedDesign}

\begin{figure}[hbt!]
 \centering
 \includegraphics[width=0.75\textwidth]{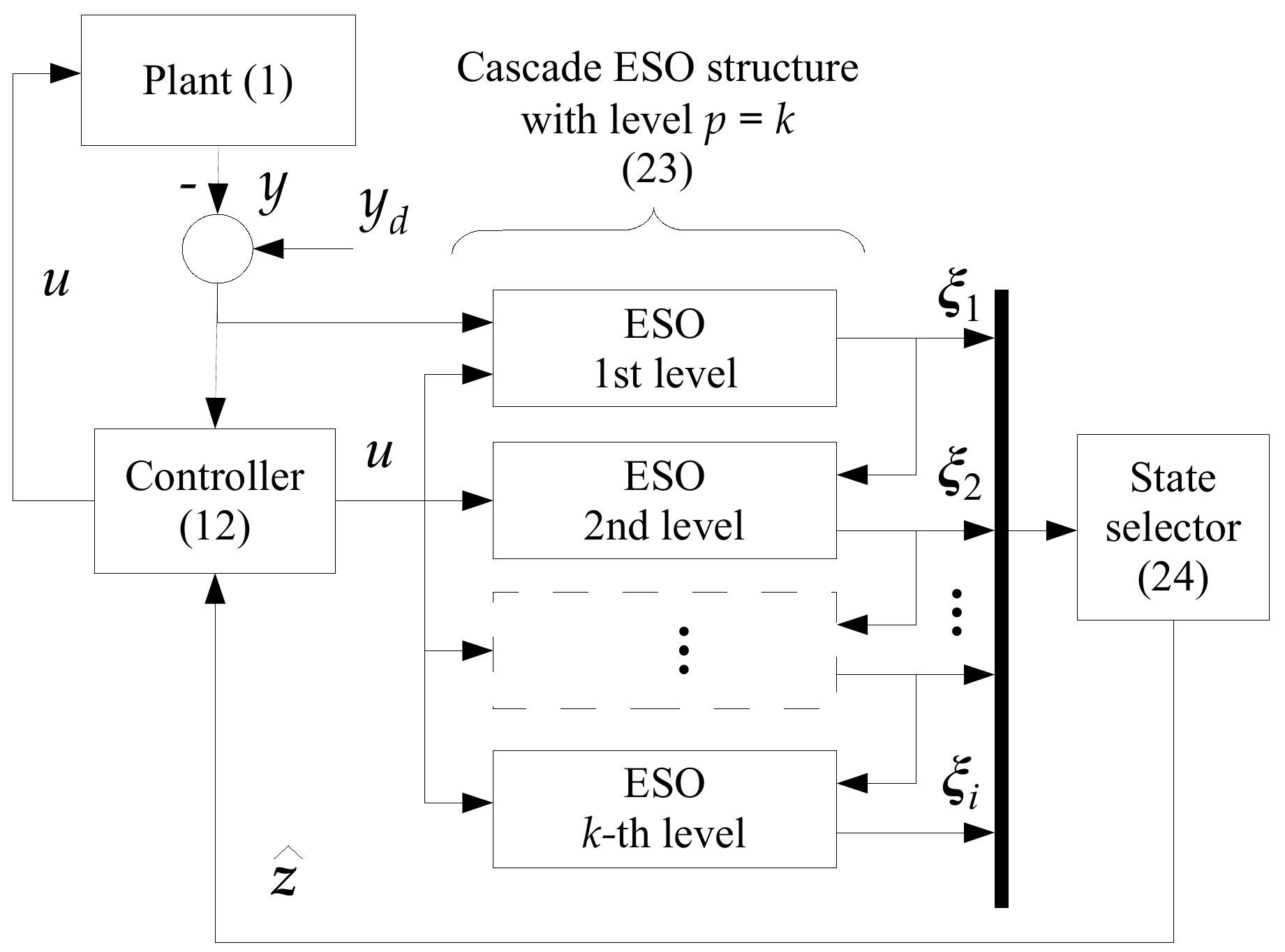}
 \caption{{Block diagram of an ADRC structure including proposed cascade ESO for $p=k$.}}
 \label{fig:BlockDiagramISATgeneralCascade}
\end{figure}

In the general case, we may continue the procedure of estimating  residual observation errors by increasing the level of observer cascade to the arbitrarily chosen value $p=k$ such that $k\in\integerNumbers$ and $k\geq2$ . The cascade observer with $i$ levels can be written down as
\begin{align}
    \observerStageStateDerivative{1}(t)&=\stateMatrix{n+1}\observerStageState{1}(t) + \inputVectorExtendedState{n+1}\controlSignalFieldEstimate\controlSignal(t)+\observerGainVector{1,n+1}(\systemOutput(t)-\outputVector{n+1}^\top\observerStageState{1}(t)), \nonumber \\
    \observerStageStateDerivative{i}(t) &= \stateMatrix{n+1}\observerStageState{i}(t) + \inputVectorExtendedState{n+1}\left(\controlSignalFieldEstimate\controlSignal(t)+\inputVector{n+1}^\top\sum_{j=1}^{i-1}\observerStageState{j}(t)\right) \nonumber \\
    &+\observerGainVector{i,n+1}\outputVector{n+1}^\top(\observerStageState{i-1}(t)-\observerStageState{i}(t)),
    \label{eq:iStageObserver}
\end{align}
for $i\in\{2,...,p\}$ and $\observerStageBandwidth{i}=\observerBandwidthMultiplier\observerStageBandwidth{i-1}$ for $\observerBandwidthMultiplier>1$. The estimate of total disturbance implemented in \eqref{eq:controlSignalSimple} can be taken from the new extended state estimate expressed as
\begin{align}
    \extendedStateEstimate := \observerStageState{i} + \inputVector{n+1}\inputVector{n+1}^\top\sum_{j=1}^{i-1}\observerStageState{j}.
\end{align}
{A block diagram of the ADRC control structure with the proposed cascade ESO (in its general form $p=k$) for the system \eqref{eq:systemModel} is shown in Fig.~\ref{fig:BlockDiagramISATgeneralCascade}.}

The observation error for the $i$-level cascade ESO is defined as
%
\begin{align}
    \observerStageObservationError{i} \triangleq \extendedState - \observerStageState{i} - \inputVector{n+1}\inputVector{n+1}^\top\sum_{j=1}^{i-1}\observerStageState{j},
\end{align}
with the dynamics expressed (after some algebraic manipulations made according to \eqref{eq:extendedStateSystem}, and \eqref{eq:iStageObserver}) as
%
\begin{align}
    \observerStageObservationErrorDerivative{i}
    &= \observationErrorStateMatrix{i}\observerStageObservationError{i} + {\residualObservationErrorMatrix{i,i-1}}\observerStageObservationError{i-1} +\inputVector{n+1}^\top\totalDisturbanceDerivative+\sum_{j=1}^{i-2}\residualObservationErrorMatrix{i,j}\observerStageObservationError{j}+\measurementNoiseVector{i}\measurementNoise,
\end{align}
for $\observationErrorStateMatrix{i} = \stateMatrix{n+1}-\observerGainVector{i,n+1}\outputVector{n+1}^\top$, $\residualObservationErrorMatrix{i,i-1} = \observerGainVector{i,n+1}\outputVector{n+1}^\top-\inputVector{n+1}\inputVector{n+1}^\top\observerGainVector{i-1,n+1}\outputVector{n+1}^\top$, $\residualObservationErrorMatrix{i,j} = -\inputVector{n+1}\inputVector{n+1}^\top\observerGainVector{j,n+1}\outputVector{n+1}^\top$, and $ \measurementNoiseVector{i} = -\inputVector{n+1}\inputVector{n+1}^\top\observerGainVector{1,n+1}\color{black}$.
Using the linear change of coordinates  $\observerStageState{i}=\stateTransformationMatrix{i}\transformedState{i}$ for $\stateTransformationMatrix{i}\triangleq\textrm{diag}\{\color{black}-\observerStageBandwidth{i}^{-n},...,-\observerStageBandwidth{i}^{-1}\color{black}
,1\}\in\realNumbers^{n+1\times n+1}$ and $\transformedState{i}\in\realNumbers^{n+1}$, we may rewrite the  observation error dynamics concerning $i$-th level structure as
\begin{align}
    \transformedStateDerivative{i} &= \underbrace{\stateTransformationMatrix{i}^{-1}\observationErrorStateMatrix{i}\stateTransformationMatrix{i}}_{\observerStageBandwidth{i}\transformedStateStateMatrix{i}}\transformedState{i}+\stateTransformationMatrix{i}^{-1}\residualObservationErrorMatrix{i,i-1}\stateTransformationMatrix{i-1}\transformedState{i-1}+\stateTransformationMatrix{i}^{-1}\inputVector{n+1}\totalDisturbanceDerivative \nonumber \\
    &+\stateTransformationMatrix{i}^{-1}\sum_{j=1}^{i-2}\residualObservationErrorMatrix{i,j}\stateTransformationMatrix{j}\transformedState{j} + \stateTransformationMatrix{i}^{-1}\measurementNoiseVector{i}\measurementNoise .
    \label{eq:transformedIthState}
\end{align}

To prove the boundedness of observation errors in the general case of cascade ESO, let us propose a Lyapunov function candidate defined as
$\observerStageLyapunovFunction{i} \triangleq \transformedState{i}^\top\lyapunovEquationSolutionMatrix{i}\transformedState{i}, \ \observerStageLyapunovFunction{i}:\realNumbers^{n+1}\color{black}\rightarrow\realNumbers$ and bounded by  $\minEigenvalue{\lyapunovEquationSolutionMatrix{2}}\module{\transformedState{2}}\leq\observerStageLyapunovFunction{2}\leq\maxEigenvalue{\lyapunovEquationSolutionMatrix{2}}\module{\transformedState{2}}$,
where $\lyapunovEquationSolutionMatrix{i}$ is a solution of an algebraic Lyapunov equation \eqref{eq:lyapunovEquation}. The derivative of $\observerStageLyapunovFunction{i}$ is calculated upon \eqref{eq:transformedIthState} and takes the form
\begin{align}
    \observerStageLyapunovFunctionDerivative{i} &= -\observerStageBandwidth{i}\transformedState{i}^\top\transformedState{i}+ 2\transformedState{i}^\top\lyapunovEquationSolutionMatrix{i}\stateTransformationMatrix{i}^{-1}\measurementNoiseVector{i}\measurementNoise + 2\transformedState{i}^\top\lyapunovEquationSolutionMatrix{i}\stateTransformationMatrix{i}^{-1}\inputVector{n+1}\totalDisturbanceDerivative \nonumber \\
    &+2\transformedState{i}^\top\lyapunovEquationSolutionMatrix{i}\stateTransformationMatrix{i}^{-1}\residualObservationErrorMatrix{i,i-1}\stateTransformationMatrix{i-1}\transformedState{i-1}+2\transformedState{i}^\top\lyapunovEquationSolutionMatrix{i}\stateTransformationMatrix{i}^{-1}\sum_{j=1}^{i-2}\residualObservationErrorMatrix{i,j}\stateTransformationMatrix{j}\transformedState{j} \nonumber \\
    &\leq -\observerStageBandwidth{i}\module{\transformedState{i}}^2 + 2\module{\lyapunovEquationSolutionMatrix{i}}\module{\transformedState{i}}\maximalGainCoefficient\color{black}\observerStageBandwidth{1}^{n+1}\abs{\measurementNoise} +2\module{\lyapunovEquationSolutionMatrix{i}}\module{\transformedState{i}}\abs{\totalDisturbanceDerivative}  \nonumber \\
    &+ 2\frac{\maximalGainCoefficient\observerStageBandwidth{i}^{n+1}}{\observerStageBandwidth{i-1}^{n}}\module{\lyapunovEquationSolutionMatrix{i}}\module{\transformedState{i-1}}\module{\transformedState{i}}+2\module{\lyapunovEquationSolutionMatrix{i}}\module{\transformedState{i}}\sum_{j=1}^{i-2}\observerStageBandwidth{j}\module{\transformedState{j}},
\end{align}
that holds
\begin{align}
    \observerStageLyapunovFunctionDerivative{i} &\leq -\observerStageBandwidth{i}(1-\observerStageMajorizationConstant{i})\module{\transformedState{i}}^2 \ \textrm{for} \nonumber \\
    \module{\transformedState{i}}&\geq \frac{2\maximalGainCoefficient\observerStageBandwidth{i}^{n}\module{\lyapunovEquationSolutionMatrix{i}}}{\observerStageMajorizationConstant{i}\observerStageBandwidth{i-1}^{n}}\module{\transformedState{i-1}} \nonumber \\
    &+\frac{2\module{\lyapunovEquationSolutionMatrix{i}}}{\observerStageMajorizationConstant{i}\observerStageBandwidth{i}}\sum_{j=1}^{i-2}\observerStageBandwidth{j}\module{\transformedState{j}} + \frac{2\module{\lyapunovEquationSolutionMatrix{i}}}{\observerStageMajorizationConstant{i}\observerStageBandwidth{i}}\abs{\totalDisturbanceDerivative} + \frac{2\module{\lyapunovEquationSolutionMatrix{i}}\maximalGainCoefficient\color{black}\observerStageBandwidth{1}^{n+1}}{\observerStageMajorizationConstant{i}\observerStageBandwidth{i}}\abs{\measurementNoise} \nonumber \\
    &= \frac{2\maximalGainCoefficient\observerBandwidthMultiplier^{n}\module{\lyapunovEquationSolutionMatrix{i}}}{\observerStageMajorizationConstant{i}}\module{\transformedState{i-1}}+ \frac{2\module{\lyapunovEquationSolutionMatrix{i}}}{\observerStageMajorizationConstant{i}}\sum_{j=1}^{i-2}\frac{1}{\observerBandwidthMultiplier^{i-j}}\module{\transformedState{j}} \nonumber \\
    &+\frac{2\module{\lyapunovEquationSolutionMatrix{i}}}{\observerStageMajorizationConstant{i}\observerBandwidthMultiplier^{i-1}\observerStageBandwidth{1}}\abs{\totalDisturbanceDerivative} + \frac{2\module{\lyapunovEquationSolutionMatrix{i}}\maximalGainCoefficient\color{black}\observerStageBandwidth{1}^{n}}{\observerStageMajorizationConstant{i}\observerBandwidthMultiplier^{i-1}}\abs{\measurementNoise}.
\end{align}
According to the ISS property, steady-state values of the transformed observation error are limited by
\begin{align}
    \limsup_{t\rightarrow\infty}\module{\transformedState{i}(t)}&\leq \issFactor\frac{2\maximalGainCoefficient\observerBandwidthMultiplier^{n}\module{\lyapunovEquationSolutionMatrix{i}}}{\observerStageMajorizationConstant{i}}\limsup_{t\rightarrow\infty}\module{\transformedState{i-1}(t)} \nonumber \\
    &+ \issFactor\frac{2\module{\lyapunovEquationSolutionMatrix{i}}}{\observerStageMajorizationConstant{i}}\sum_{j=1}^{i-2}\frac{1}{\observerBandwidthMultiplier^{i-j}}\limsup_{t\rightarrow\infty}\module{\transformedState{j}(t)} \nonumber \\
    &+\issFactor\frac{2\module{\lyapunovEquationSolutionMatrix{i}}}{\observerStageMajorizationConstant{i}\observerBandwidthMultiplier^{i-1}\observerStageBandwidth{1}}\sup_{t\geq0}\abs{\totalDisturbanceDerivative(t)} + \issFactor\frac{2\module{\lyapunovEquationSolutionMatrix{i}}\maximalGainCoefficient\color{black}\observerStageBandwidth{1}^{n-1}}{\observerStageMajorizationConstant{i}\observerBandwidthMultiplier^{i-1}}\sup_{t\geq0}\abs{\measurementNoise(t)},
    \label{eq:finalResult}
\end{align}
where $\issFactor$ is taken from \eqref{eq:firstStageCascadeResult}.
Due to the recursive character of the obtained asymptotic relation and the result \eqref{eq:firstStageCascadeResult}, we may also write that $\limsup_{t\rightarrow\infty}\module{\transformedState{i}}\leq c_1\totalDisturbanceDerivativeUpperBound + c_2 \measurementNoiseUpperBound$ for some positive constants $c_1, c_2$ dependent on the parameters $\observerStageBandwidth{1}$ and $\observerBandwidthMultiplier$. As it was presented for previously described observer structures, transformed observation error is generally bounded, $\module{\transformedState{i}}\rightarrow0$ as $\observerStageBandwidth{1}\rightarrow\infty$ and $t\rightarrow\infty$ when $\measurementNoise(t)\equiv0$, and $\limsup_{t\rightarrow\infty}\module{\transformedState{i}(t)}=0$ when both $\measurementNoise(t)\equiv0$ and $\totalDisturbanceDerivative(t)\equiv0$.
Similarly to the issue described in Remark~\ref{rem:3} for the single-level observer structure, the original observation error for the $i$-level cascade observer  $\module{\observerStageObservationError{i}}\leq\maxEigenvalue{\stateTransformationMatrix{i}}\module{\transformedState{i}}=\max\left\{\frac{1}{\observerStageBandwidth{i}^{-n}},1\right\}\module{\transformedState{i}}$ and thus holds the ISS property itself. \color{black} The obtained result presents the practical stability achieved for the non-zero measurement noise and total disturbance derivative which are the perturbations of the analyzed system.\color{black}



\section{Numerical verification}

\subsection{Generic example}
\label{ref:genericEx}

\paragraph{Methodology}


In order to evaluate the effectiveness of the proposed cascade ESO-based ADRC design (Sect.~\ref{sec:concept}) in terms of control objective realization and measurement noise attenuation, its results are quantitatively compared with the results from a standard, single ESO-based ADRC design (Sect.~\ref{sec:ConventionalESO}). For the purpose of this case study, a following second order \color{black} time-varying single-input single-output \color{black} system in form of~\eqref{eq:systemModel} is considered:
\begin{align}
    \begin{cases}
        \stateVectorDerivative(t) = \stateMatrix{2}\stateVector(t)+\inputVector{2}(\systemNonlinearDynamics(\stateVector)+\color{black}g(\stateVector,t)\color{black}\controlSignal(t)+\disturbance(t)), \\
        \systemOutput(t) = \outputVector{2}^\top\stateVector(t)+\measurementNoise(t),
    \end{cases}
    \label{eq:rewrittenSystemDynamicsSpecCase}
\end{align}
with $\stateVector{}=\begin{bsmallmatrix}
  x_1 \\
  x_2
\end{bsmallmatrix}$, $\systemNonlinearDynamics(\stateVector{}) = -\stateVectorElement{1}-2\stateVectorElement{2}$, and \color{black}$g(\stateVector,t)=\frac{1+0.2\tanh(t-2)}{\abs{\stateVectorElement{1}}+1}$\color{black}. The core linear dynamics of~\eqref{eq:rewrittenSystemDynamicsSpecCase} is affected (starting at $t=3.5$s) by an external, nonlinear disturbance $\disturbance(t)=5\sin(9 t)$. The control objective is to make the output $y(t)$ track a desired trajectory $y_d(t)$ in the presence of measurement noise $w(t)$ and despite the influence of $\disturbance(t)$. Signal $y_d(t)$ here is a step function, additionally filtered by a stable dynamics $G_f(s)=1/(0.1s + 1)^5$ to minimize the observer peaking. The control action $u$ is defined the same for all tested cases as in~\eqref{eq:controlSignalSimple}, with $v=\ddot{y}_d+4(\dot{y}_d-\observerStageStateElement{p,2}{})+4(y_d-\observerStageStateElement{p,1})$. \color{black} The utilized control design is a well-known structure, studied in details for example in~\cite{Gao-stabilitySingPert}\color{black}. To keep the conciseness of notation, we treat the standard ESO as the case with $p=1$. Two comparison tests are performed.





\paragraph{Sim1A: trajectory tracking without measurement noise} In order to establish a fair base for further comparison in the presence of measurement noise, similar performance between the tested control designs is achieved first in the idealized, noise-less conditions ($w(t)=0$). Since total disturbance estimation is the key element in any ADRC approach, the focus here is on providing similar reconstruction quality of $\hat{d}$ in terms of minimizing integral criterion $\int_{t_0}^T|\tilde{z}_3(t)|dt$, with $\tilde{z}_3(t)$ being the total disturbance observation error taken from $\extendedStateObservationError=[\tilde{z}_1 \ \tilde{z}_2 \ \tilde{z}_3]^\top\triangleq\observerStageObservationError{p}$, \textcolor{black}{$t_0=0$s being the integration start time and $T=5$s being its finish time}. Since the tested control algorithms do not share similar observer structure, nor have the same number of tuning parameters, it is arbitrarily decided to use a tuning methodology of single observer bandwidth parameterization from~\cite{Gao-pole-placement}, which is based on a standard pole-placement procedure (cf.~Remark~\ref{remPP}). The heuristically selected observer bandwidths from Table~\ref{tab:the_tableSim1} provide similar total disturbance reconstruction quality among the tested algorithms, as confirmed by Table~\ref{tab:qualityCriteriaSim1}. Two integral criteria are additionally introduced to evaluate overall tracking accuracy ($10^3\int_{t_0}^T|e(t)|dt$) and energy usage ($\int_{t_0}^Tu^2(t)dt$){, with $e(t)\triangleq y_d(t)-y(t)$ being the feedback error}.


\begin{table}[p]
    \centering
    \caption{\textcolor{black}{Observer structures and parameters used in the comparison in Sim1A and Sim1B}.}
    \label{tab:the_tableSim1}
    \renewcommand{\arraystretch}{1}
        \begin{tabular}{|c|c|c|}
        \hline
        Observer type & Observer bandwidth\\
        \hline\hline
        Standard ESO $p$=1 & $\omega_{o1}=250$\\
        \hline
        Proposed ESO $p$=2 & $\omega_{o1}=60.82$, $\omega_{o2}=2\omega_{o1}$\\
        \hline
        Proposed ESO $p$=3 & $\omega_{o1}=33.98$, $\omega_{o2}=2\omega_{o1}$, $\omega_{o3}=2\omega_{o2}$\\
        \hline
    \end{tabular}
    \renewcommand{\arraystretch}{1}
\end{table}

\begin{table}[p]
 \centering
 \caption{\textcolor{black}{Assessment based on selected integral quality criteria in Sim1A and Sim1B.}}
\label{tab:qualityCriteriaSim1}
\begin{tabular}{|c|c|c|c|c|}
\cline{1-5}
\multirow{2}{*}{\rot{Test~}} & \multirow{2}{*}{Observer type} & \multicolumn{3}{c|}{Criterion} \\ \cline{3-5}
 &  & $10^3\int_{t_0}^T|e(t)|dt$ & $\int_{t_0}^T u^2(t)dt$ & $\int_{t_0}^T|\tilde{z}_3(t)|dt$ \\ \hline\hline
\multicolumn{1}{|c|}{\multirow{3}{*}{\rot{\textit{Sim1A}}}} & Standard ESO $p$=1 & 19.525 & 79.076 & \textbf{0.956} \\ \cline{2-5}
\multicolumn{1}{|c|}{} & Proposed ESO $p$=2 & 12.351 & 89.958 & \textbf{0.957} \\ \cline{2-5}
\multicolumn{1}{|c|}{} & Proposed ESO $p$=3 & 8.346 & 84.804 & \textbf{0.955} \\ \hline
\multicolumn{1}{|c|}{\multirow{3}{*}{\rot{\textit{Sim1B}}}} & Standard ESO $p$=1 & 19.738 & 193.93 & 8.785 \\ \cline{2-5}
\multicolumn{1}{|c|}{} & Proposed ESO $p$=2 & 12.833 & 100.61 & 3.162 \\ \cline{2-5}
\multicolumn{1}{|c|}{} & Proposed ESO $p$=3 & 8.952 & 93.67 & 3.006 \\ \hline
\end{tabular}
\end{table}

\paragraph{Sim1B: trajectory tracking with measurement noise} This time, a more realistic scenario is considered in which the output measurement noise is present (i.e. $w(t)\neq 0$), hence the only change with respect to test \textit{Sim1A} is the introduction of the band-limited white noise $w(t)$ with power \color{black} $1e^{-9}$. The sampling frequency is set to $100$Hz.\color{black} 

\paragraph{Results}

\begin{figure}[t]
  \centering
    \includegraphics[width=\textwidth]{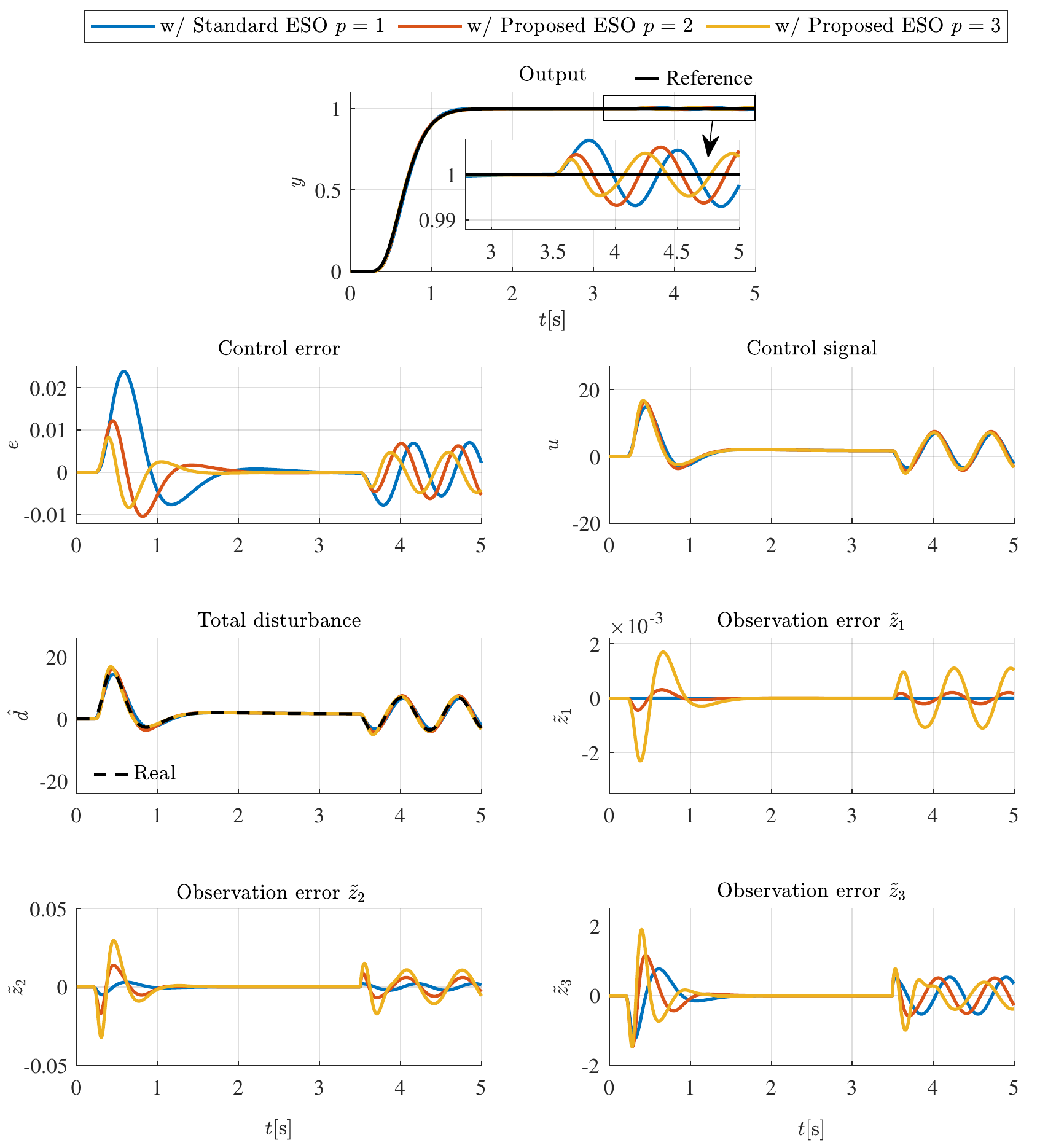}
    \caption{\textcolor{black}{Results of test \textit{Sim1A} (i.e. without measurement noise: $w(t)=0$).}}
    \label{fig:SimA}
\end{figure}

The results of \textit{Sim1A} are gathered in Fig.~\ref{fig:SimA}. The selected observer bandwidths provide visually comparable results of total disturbance estimation error $\tilde{z}_3(t)$ and energy consumption $u(t)$ in all tested control designs, even though both cascade ESO-based designs have significantly lower observer bandwidths than standard approach. The proposed cascade ESO structures also improve the tracking quality $e(t)$ despite having lower observer bandwidths. The observation error $\tilde{z}_1(t)$ differs among tested techniques noticeably, however, it is expected result since in the cascade observer design, the importance of output signal estimation is being shifted and favors total disturbance estimation instead. It is done deliberately and has practical justification as it is often the case that the output signal is available directly through measurement, hence its close estimation would be redundant.



\begin{figure}[t]
  \centering
    \includegraphics[width=\textwidth]{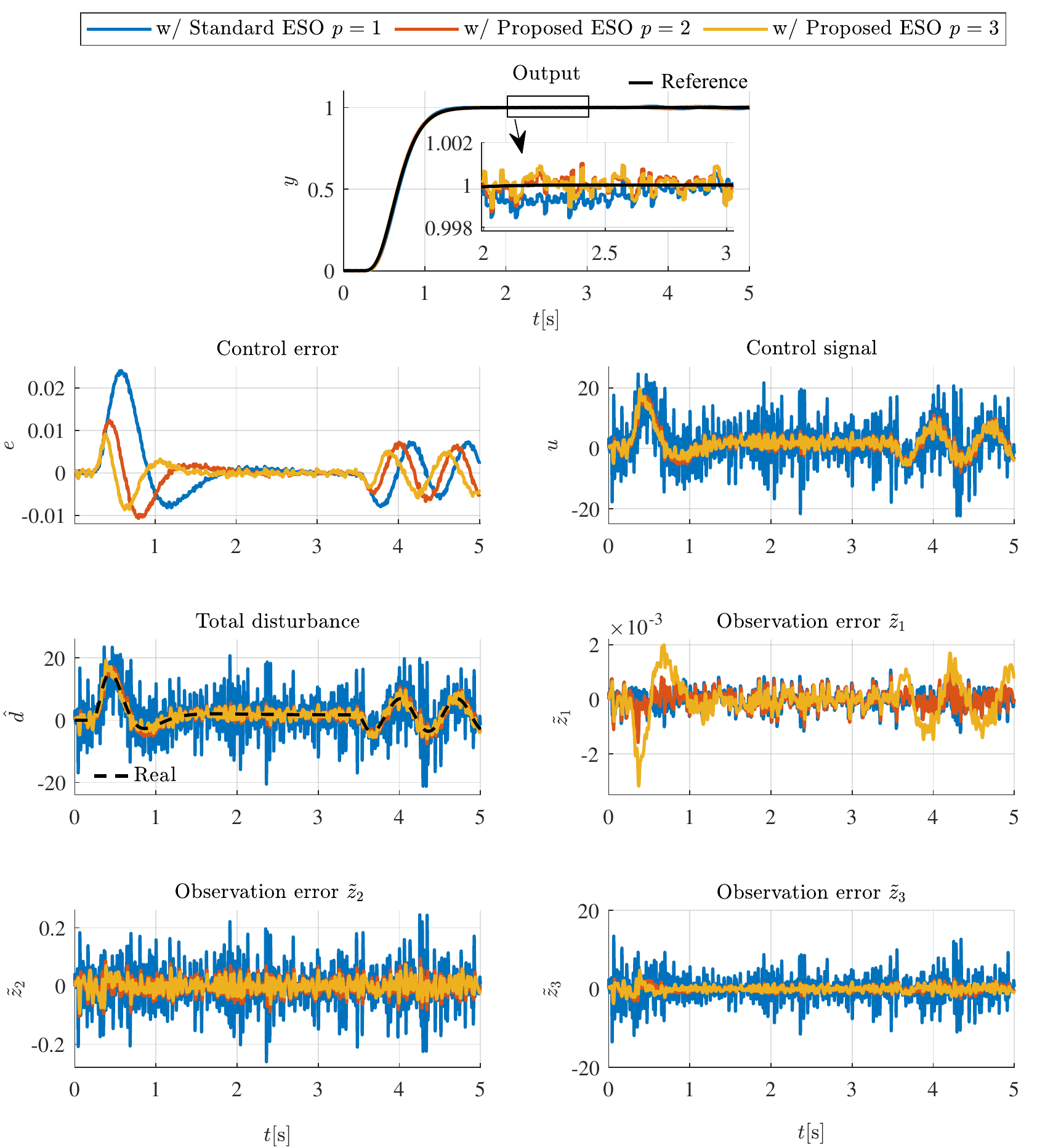}
    \caption{\textcolor{black}{Results of test \textit{Sim1B} (i.e. with measurement noise: $w(t)\neq 0$).}}
    \label{fig:SimB}
\end{figure}

The results of \textit{Sim1B} are gathered in Fig.~\ref{fig:SimB}. \color{black}One can see that despite the presence of the measurement noise visible on zooms embedded in the plot presenting the output signal\color{black}, the cascade ESO provides improvement in the noise attenuation over standard ESO, especially in the quality of total disturbance reconstruction and, consequently, in the control signal profile (Table~\ref{tab:qualityCriteriaSim1}). With the extra layer of cascade ($p=3$), the amplitude of noise is further reduced. It is not surprising since with the increase of cascade level, lower observer bandwidths can be selected for estimating signals contaminated with high-frequency noise (hence noise is not over-amplified). At the same time, high observer bandwidths can be selected for estimating signals which are affected by the noise to a smaller extent, hence high tracking capabilities can be retained.

\begin{figure}[p]
  \thisfloatpagestyle{empty}
  \centering
    \includegraphics[width=0.49\textwidth]{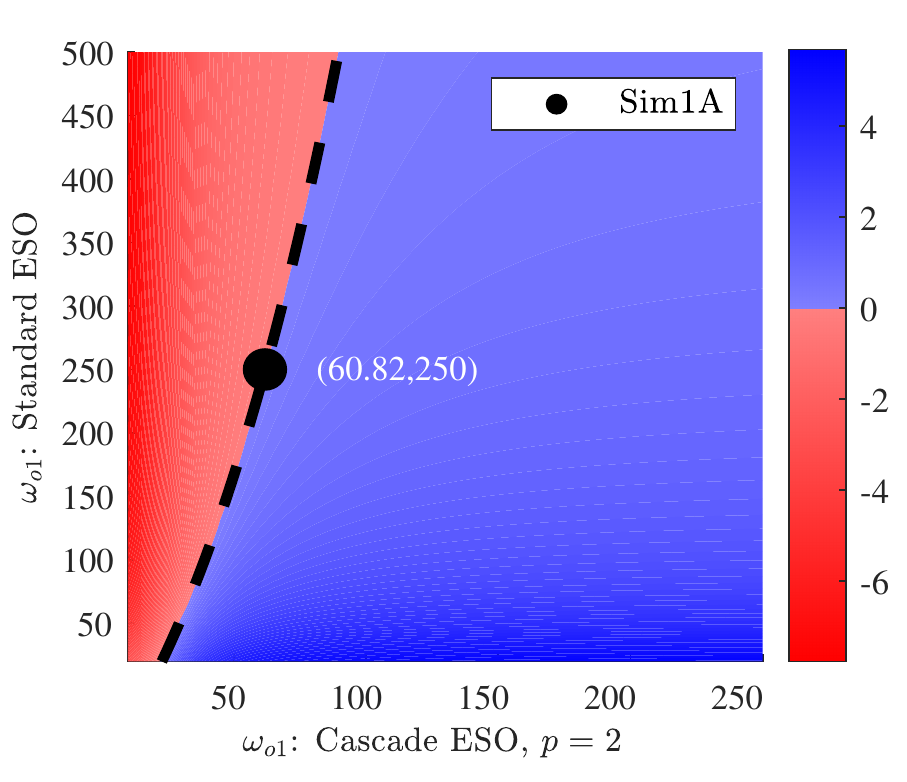}
    \includegraphics[width=0.49\textwidth]{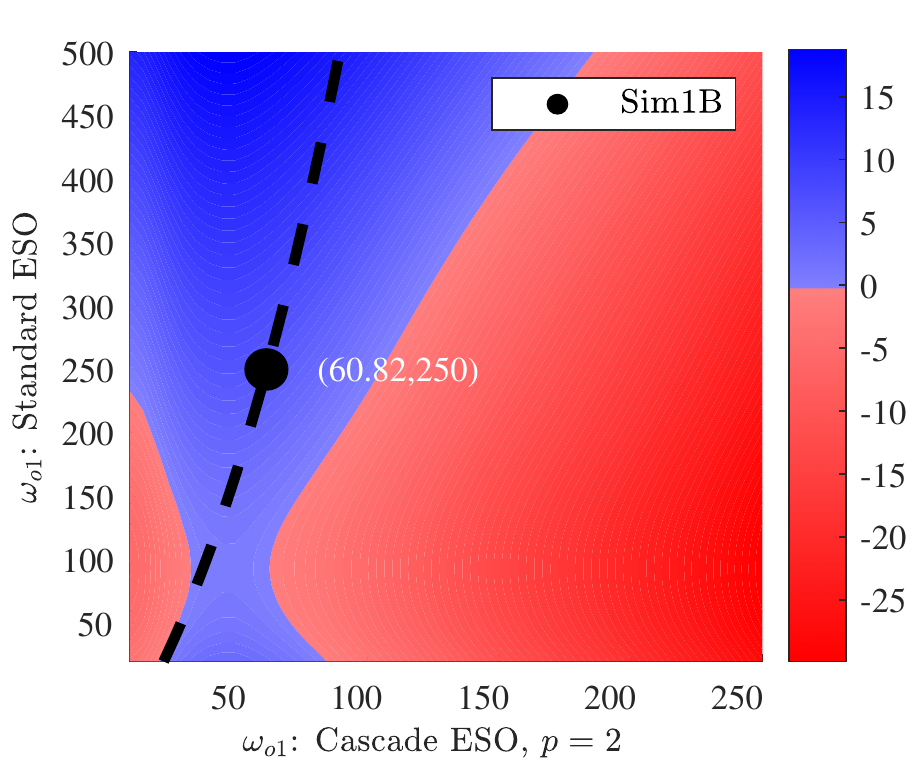} \\
    \includegraphics[width=0.49\textwidth]{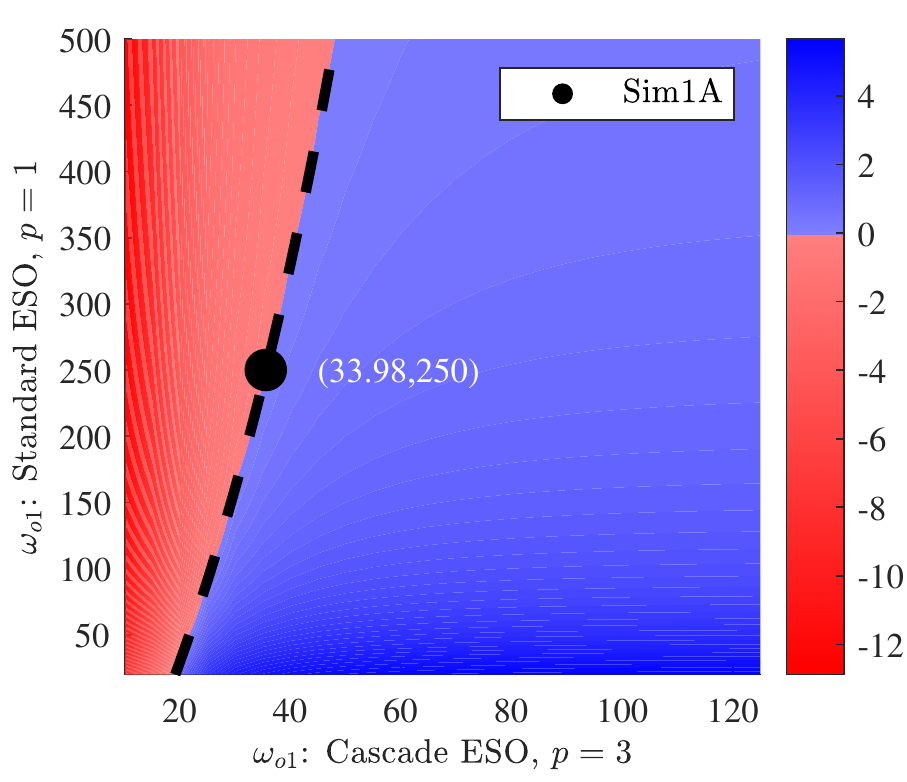}
    \includegraphics[width=0.49\textwidth]{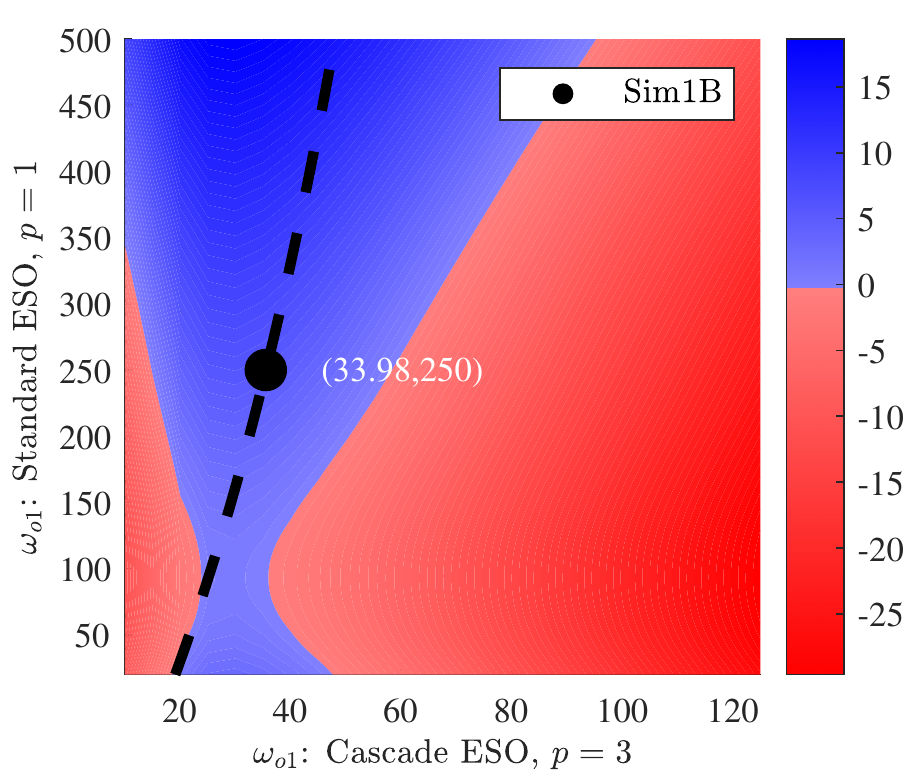} \\
    \includegraphics[width=0.49\textwidth]{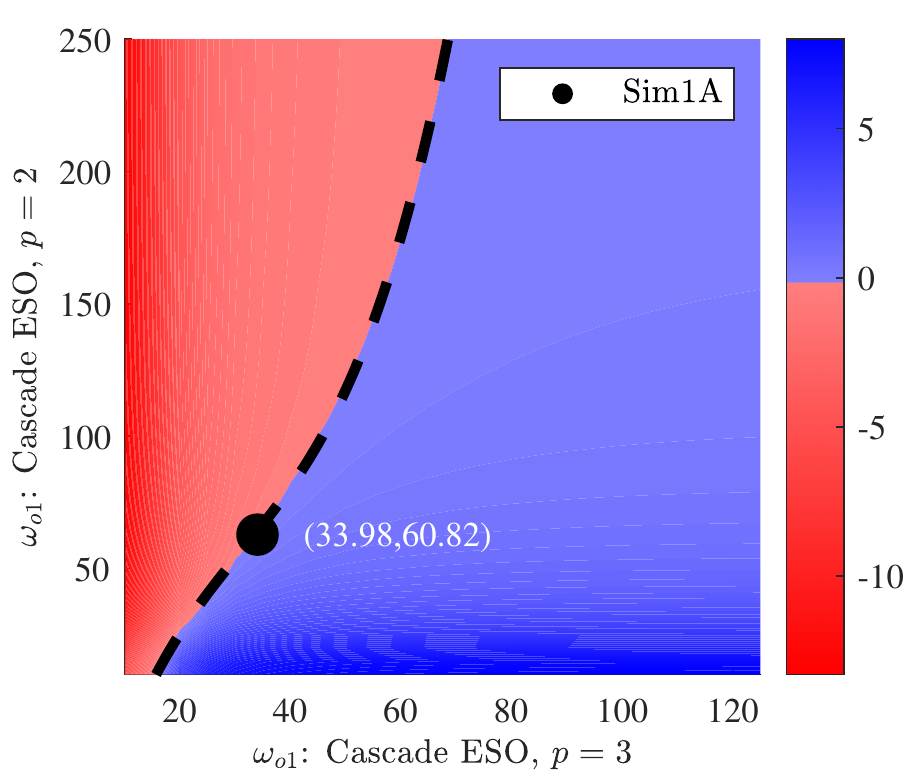}
    \includegraphics[width=0.49\textwidth]{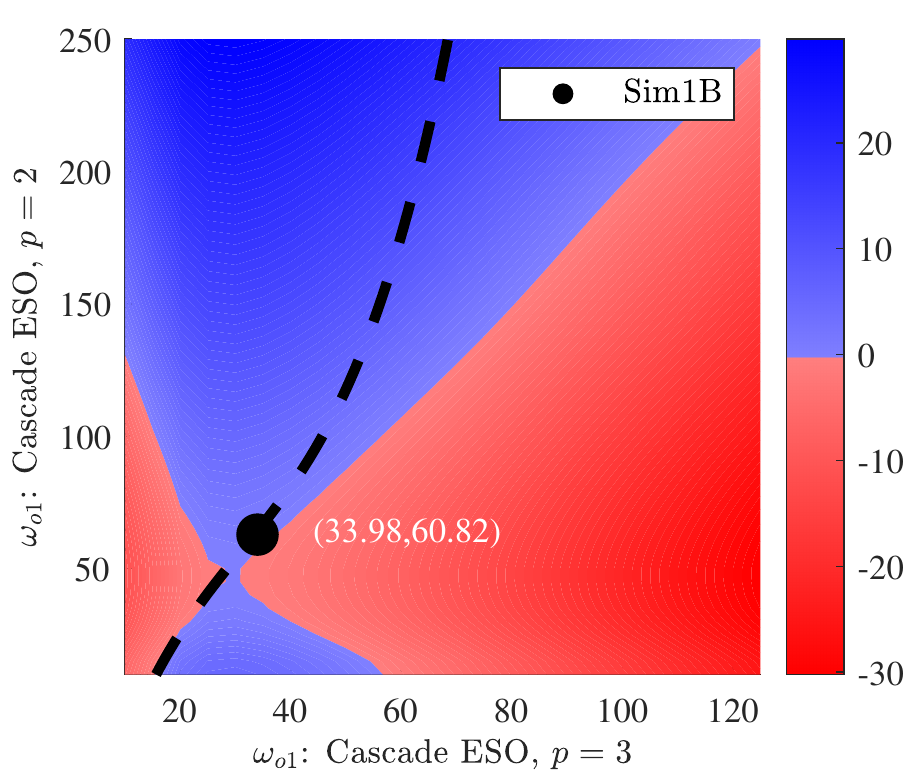}
    \caption{\textcolor{black}{Difference between $\int_{t_0}^T|\tilde{z}_3(t)|dt$ values obtained for the standard and the proposed ESOs ($p=2$ and $p=3$). The left-hand side figures are related to the simulations without measurement noise ($w(t)=0$), while the figures on the right-hand side are associated with the cases where the measurement noise is present ($w(t)\neq0$). Black dots represent the specific pair of observer bandwidths used in both tests (see Table~\ref{tab:the_tableSim1}).}}
    \label{fig:alphaBetaComparison}
\end{figure}

Additionally, Fig.~\ref{fig:alphaBetaComparison} shows a relative quality of a total disturbance estimation, understood as the difference of $\int_{t_0}^T|\tilde{z}_3(t)|dt$ criteria between the standard and proposed ESO-based designs in tests \textit{Sim1A} (left) and \textit{Sim1B} (right).
The blue color represents area where the total disturbance is estimated more precisely with ESO associated with the horizontal axis, while the red color represents the area where an ESO structure related to the vertical axis provides better estimation.
The dashed lines in the left-hand side figures connect points with equal values of estimation precision, i.e., cases having similar convergence speed of observation error. Black dots placed on each plot are representing the parameter values utilized in the simulations illustrated in time-domain in Fig.~\ref{fig:SimA} and Fig.~\ref{fig:SimB}.  When the same dashed line is plotted in the right-hand side figure associated with the comparison of standard ESO and the proposed one (both for $p=2$ and $p=3$), it is surrounded by blue area for high observer bandwidths ($\observerStageBandwidth{1}$) of standard ESO. This confirms numerically that the use of cascade ESO is justified when high observer gains are demanded, as it can improve the estimation precision (connected with noise attenuation) for similar convergence speed. The comparison of cascaded ESO structures for $p=2$ and $p=3$ shows that in the presence of measurement noise, the estimation quality is (a) equal for both structures, when the dashed line lays on the border of blue and red area, (b) better for $p=2$ level structure for a small range of relatively low values of $\observerStageBandwidth{1}$ when the dashed line is on the red area, and (c) more precise for $p=3$ level cascade ESO in the case of dashed line laying on the blue area. However, case with $p=3$ revealed (expectedly) that there is a limit to the cascade level increase. And although it is shown through the obtained numerical results that the introduction of cascade ESO structure can be beneficial in sensor noise suppression, both the specific value of $p$ and the ESO gains should be tailored on a case-by-case basis. General methodology for $p$ selection and observer tuning is still an open question.

\color{black}

\color{black}

\subsection{PVTOL aircraft example}

\paragraph{Methodology}

In relation to the generic example presented in Sect.~\ref{ref:genericEx}, this time, a more realistic model of a planar vertical take-off and landing (PVTOL) aircraft is considered for the verification of the proposed cascade ESO-based ADRC. According to~\cite{aguilar2018}, a PVTOL aircraft dynamics can be well approximated with a simplified and normalized model
\begin{align}
    \pvtolXSecondDerivative &= -\sin(\pvtolOrientation)\left(\pvtolThrust+\pvtolWindThrust\right), \nonumber \\
    \pvtolYSecondDerivative &= \cos(\pvtolOrientation)\left(\pvtolThrust+\pvtolWindThrust\right)-1, \label{eq:pvtolDynamics} \\
    \pvtolOrientationSecondDerivative &= \pvtolMomentum + \pvtolWindMomentum, \nonumber
\end{align}
where $[\pvtolX \ \pvtolY \ \pvtolOrientation]^\top\in\realNumbers^3$ is the normalized position and orientation vector, $\pvtolThrust\in\positiveRealNumbers,\pvtolMomentum\in\realNumbers$ are the control signals corresponding to a normalized thrust and momentum of the PVTOL platform, while $\pvtolWindThrust\in\realNumbers,\pvtolWindMomentum\in\realNumbers$ are the normalized signals describing the impact of crosswind, treated here as an input-additive external disturbance. We assume here that the only measurable signals are $\pvtolX, \ \pvtolY$ and $\pvtolOrientation$.

In order to express system~\eqref{eq:pvtolDynamics} in the form of~\eqref{eq:rewrittenSystemDynamics}, dynamics of the vehicle position is expressed as a 4-th order system
\begin{align}
    \begin{bmatrix}\pvtolXFourthDerivative \\ \pvtolYFourthDerivative\end{bmatrix} &= \pvtolMatrix(\pvtolOrientation,\pvtolThrust+\pvtolWindThrust)\begin{bmatrix} \pvtolThrustSecondDerivative + \pvtolWindThrustSecondDerivative \\ \pvtolMomentum + \pvtolWindMomentumSecondDerivative \end{bmatrix}\nonumber\\
    &+ \begin{bmatrix} \sin(\pvtolOrientation)\pvtolOrientationDerivative^2(\pvtolThrust+\pvtolWindThrust)-2\cos(\pvtolOrientation)\pvtolOrientationDerivative(\pvtolThrustDerivative+\pvtolWindThrustDerivative) \\ -\cos(\pvtolOrientation)\pvtolOrientationDerivative^2(\pvtolThrust+\pvtolWindThrust)-2\sin(\pvtolOrientation)\pvtolOrientationDerivative(\pvtolThrustDerivative+\pvtolWindThrustDerivative) \end{bmatrix},
    \label{eq:pvtolExtendedModel}
\end{align}
where $\pvtolMatrix(\pvtolOrientation,\alpha) \triangleq \begin{bmatrix} -\sin(\pvtolOrientation) & -\alpha\cos(\pvtolOrientation) \\ \cos(\pvtolOrientation) & -\alpha\sin(\pvtolOrientation) \end{bmatrix}$.  \color{black} Using the controller equations developed in \cite{aguilar2018}\color{black}, defined as
\begin{align}
    \begin{bmatrix} \pvtolThrustSecondDerivative \\ \pvtolMomentum \end{bmatrix} \triangleq \pvtolMatrix^{-1}(\pvtolOrientation,\pvtolThrust)\begin{bmatrix} \pvtolVirtualControlSignalX \\ \pvtolVirtualControlSignalY \end{bmatrix},
    \label{eq:controlsPVTOL}
\end{align}
we can rewrite system~\eqref{eq:pvtolExtendedModel} as two separate subsystems with the structure given by \eqref{eq:rewrittenSystemDynamics}, i.e.,
\begin{align}
    \begin{cases}
        \stateVectorDerivative(t) = \stateMatrix{4}\stateVector(t)+\inputVector{4}\controlSignal(t)+ {\inputVector{4}\totalDisturbance(\stateVector,t)}, \\
        \systemOutput(t) = \outputVector{4}^\top\stateVector(t)+\measurementNoise(t).
    \end{cases}
    \label{eq:pvtolStateSpace}
\end{align}
For the subsystem concerning coordinate $\pvtolX$, we can specify the assignments $(\stateVector,\controlSignal,\totalDisturbance):=([\pvtolX \ \pvtolXDerivative \ \pvtolXSecondDerivative \ \pvtolXThirdDerivative]^\top,  \pvtolVirtualControlSignalX,  \pvtolTotalDisturbanceX)$, and for the coordinate $\pvtolY$, assignments $(\stateVector,\controlSignal,\totalDisturbance):=([\pvtolY \ \pvtolYDerivative \ \pvtolYSecondDerivative \ \pvtolYThirdDerivative]^\top,  \pvtolVirtualControlSignalY,  \pvtolTotalDisturbanceY)$. Now, total disturbances associated with particular subsystems  take the form of
\begin{align}
    \begin{bmatrix}\pvtolTotalDisturbanceX \\ \pvtolTotalDisturbanceY\end{bmatrix} &= \pvtolMatrix(\pvtolOrientation,\pvtolWindThrust)\begin{bmatrix} \pvtolThrustSecondDerivative \\ \pvtolMomentum \end{bmatrix} + \pvtolMatrix(\pvtolOrientation,\pvtolThrust+\pvtolWindThrust)\begin{bmatrix}  \pvtolWindThrustSecondDerivative \\  \pvtolWindMomentumSecondDerivative \end{bmatrix}  \nonumber \\
    &+ \begin{bmatrix} \sin(\pvtolOrientation)\pvtolOrientationDerivative^2(\pvtolThrust+\pvtolWindThrust)-2\cos(\pvtolOrientation)\pvtolOrientationDerivative(\pvtolThrustDerivative+\pvtolWindThrustDerivative) \\ -\cos(\pvtolOrientation)\pvtolOrientationDerivative^2(\pvtolThrust+\pvtolWindThrust)-2\sin(\pvtolOrientation)\pvtolOrientationDerivative(\pvtolThrustDerivative+\pvtolWindThrustDerivative) \end{bmatrix}.
\end{align}
The control action $u$ for each subsystem is defined for all tested cases as in~\eqref{eq:controlSignalSimple}, with $\controlSignalFieldEstimate=1$, and $v={y}_d^{(4)}+5.0625({y}_d^{(3)}-\observerStageStateElement{p,4})+13.5(\ddot{y}_d-\observerStageStateElement{p,3})+13.5(\dot{y}_d-\observerStageStateElement{p,2})+6(y_d-\observerStageStateElement{p,1})$. The desired trajectory for subsystem concerning $\pvtolX$ coordinate is $y_d(t) = \sin\left(\frac{3}{4}t\right)$, while for subsystem concerning coordinate $\pvtolY$, the desired trajectory is $y_d(t) = \frac{1}{2}\cos\left(\frac{3}{4}t\right)+1$. The initial position and orientation values are $\pvtolX(0) = 0$, $\pvtolY(0) = 0$, $\pvtolOrientation(0) = 0.6$, their derivatives are initially set to zero, while $\pvtolThrust(0)=1.5$, and $\pvtolThrustDerivative(0) = 0$. The crosswind impact is modeled as $\pvtolWindThrust = 0.3+0.2\sin(t)\cos(t)$, $\pvtolWindMomentum = 0.2 + 0.1\sin(0.5t)\cos(0.5t)$. For the sake of a clear presentation of further results, we introduce symbols $e_{i}$, $\hat{d}_i$, and $\tilde{z}_{3i}$ for $i\in\{x,z\}$ that respectively correspond to the control error $e$, the estimate of total disturbance $\hat{d}$, and the total disturbance estimation error $\tilde{z}_3$ in a particular subsystem.

Similarly to the testing procedure utilized for the generic example in Sect.~\ref{ref:genericEx}, the performance of the proposed cascade ESO-based ADRC is compared with a standard, single ESO-based ADRC. Analogously, two following tests are performed.

\paragraph{Sim2A: trajectory tracking without measurement noise}


Similar tuning procedure to the one used in Sect.~\ref{ref:genericEx} is applied to tune the observer parameters. The heuristically selected observer bandwidths (gathered in Table~\ref{tab:the_tableSim2}) provide similar total disturbance reconstruction quality among the tested algorithms, as confirmed by the values of $\int_{t_0}^T\sqrt{\tilde{z}_{3x}^2(t)+\tilde{z}_{3z}^2(t)}dt$ presented in Table~\ref{tab:qualityCriteriaSim2}. Since in the considered PVTOL aircraft control problem the initial conditions are relatively far from the reference trajectory, the integral quality indices from Table~\ref{tab:qualityCriteriaSim2} are calculated for parameters $t_0=5$s and $T=40$s to avoid integration during the initial peaking of the observer states.

\begin{table}
    \centering
    \caption{\textcolor{black}{Observer structure and parameters used in the comparison in Sim1A and Sim1B}.}
    \label{tab:the_tableSim2}
    \renewcommand{\arraystretch}{1}
        \begin{tabular}{|c|c|c|}
        \hline
        Observer type & Observer bandwidth\\
        \hline\hline
        Standard ESO $p$=1 & $\omega_{o1}=200$\\
        \hline
        Proposed ESO $p$=2 & $\omega_{o1}=25.83$, $\omega_{o2}=3\omega_{o1}$\\
        \hline
        Proposed ESO $p$=3 & $\omega_{o1}=9.65$, $\omega_{o2}=3\omega_{o1}$, $\omega_{o3}=3\omega_{o2}$\\
        \hline
    \end{tabular}
    \renewcommand{\arraystretch}{1}
\end{table}

\begin{landscape}
\begin{table}[p]
 \centering
 \caption{\textcolor{black}{Assessment based on selected integral quality criteria in Sim2A and Sim2B.}}
\label{tab:qualityCriteriaSim2}
\begin{tabular}{|c|c|c|c|c|}
\cline{1-5}
\multirow{2}{*}{\rot{Test~}} & \multirow{2}{*}{Observer type} & \multicolumn{3}{c|}{Criterion} \\ \cline{3-5}
 &  & $\int_{t_0}^T\sqrt{e_x^2(t)+e_z^2(t)}dt$ & $\int_{t_0}^T{\left[v_1^2(t)+v_2^2(t)\right]}dt$ & $\int_{t_0}^T\sqrt{\tilde{z}_{3x}^2(t)+\tilde{z}_{3z}^2(t)}dt$ \\ \hline\hline
\multicolumn{1}{|c|}{\multirow{3}{*}{\rot{\textit{Sim2A}}}} & Standard ESO $p$=1 & 3.237 & 44.508 & \textbf{0.122} \\ \cline{2-5}
\multicolumn{1}{|c|}{} & Proposed ESO $p$=2 & 3.092 & 44.525 & \textbf{0.122} \\ \cline{2-5}
\multicolumn{1}{|c|}{} & Proposed ESO $p$=3 & 2.990 & 44.552 & \textbf{0.122} \\ \hline
\multicolumn{1}{|c|}{\multirow{3}{*}{\rot{\textit{Sim2B}}}} & Standard ESO $p$=1 & 3.237 & 27813 & 1121 \\ \cline{2-5}
\multicolumn{1}{|c|}{} & Proposed ESO $p$=2 & 3.092 & 45.244 & 4.562 \\ \cline{2-5}
\multicolumn{1}{|c|}{} & Proposed ESO $p$=3 & 2.990 & 44.643 & 1.601 \\ \hline
\end{tabular}
\end{table}
\end{landscape}

\paragraph{Sim2B: trajectory tracking with measurement noise}

In this test, the only change with respect to simulation \textit{Sim2A} is an introduction of the band-limited white noise $w(t)$ with power $1e^{-17}$. The sampling sampling frequency is set to 1kHz.

\paragraph{Results}

\begin{figure}[p]
  \centering
    \includegraphics[width=0.75\textwidth]{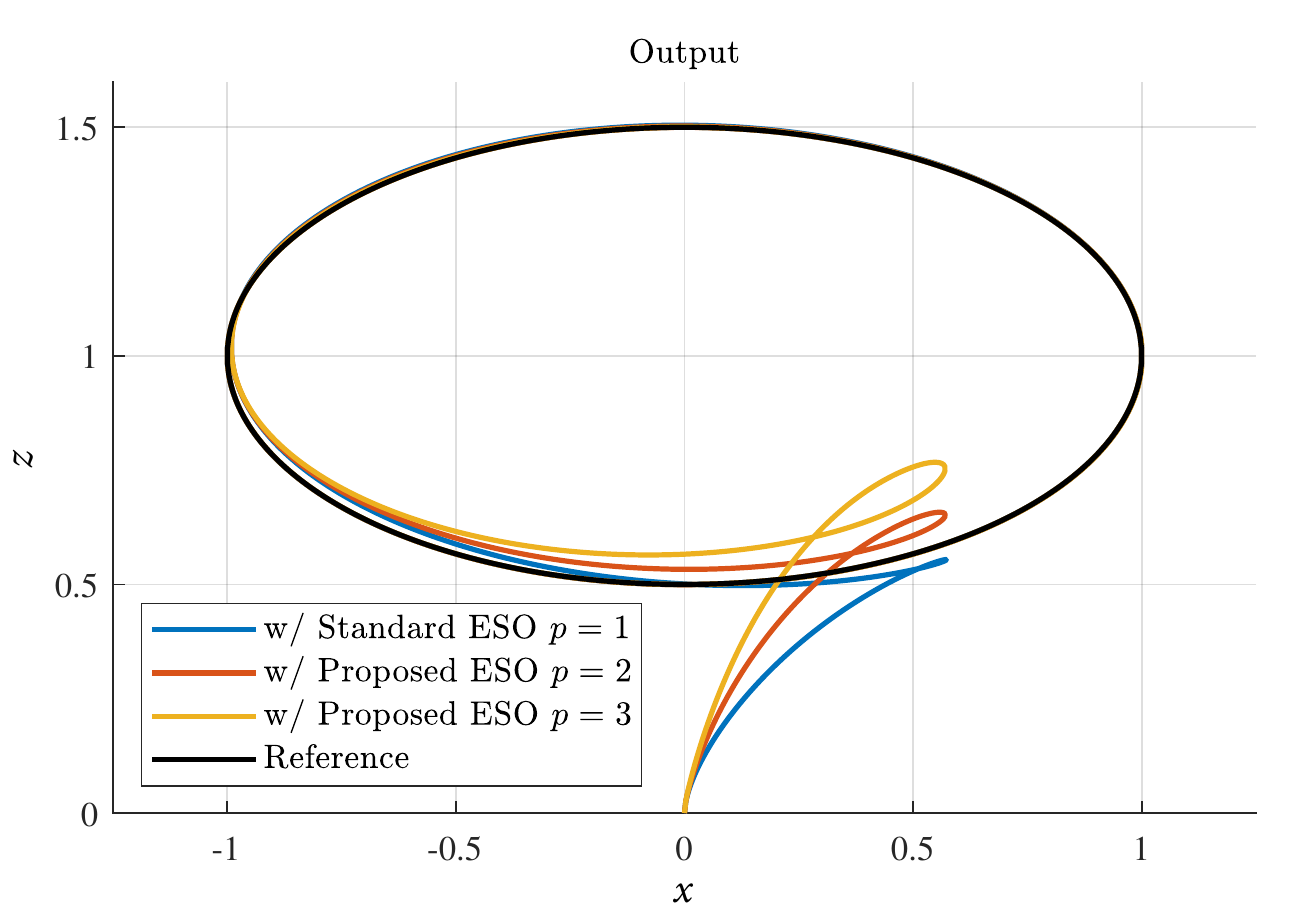}\\
    \includegraphics[width=\textwidth]{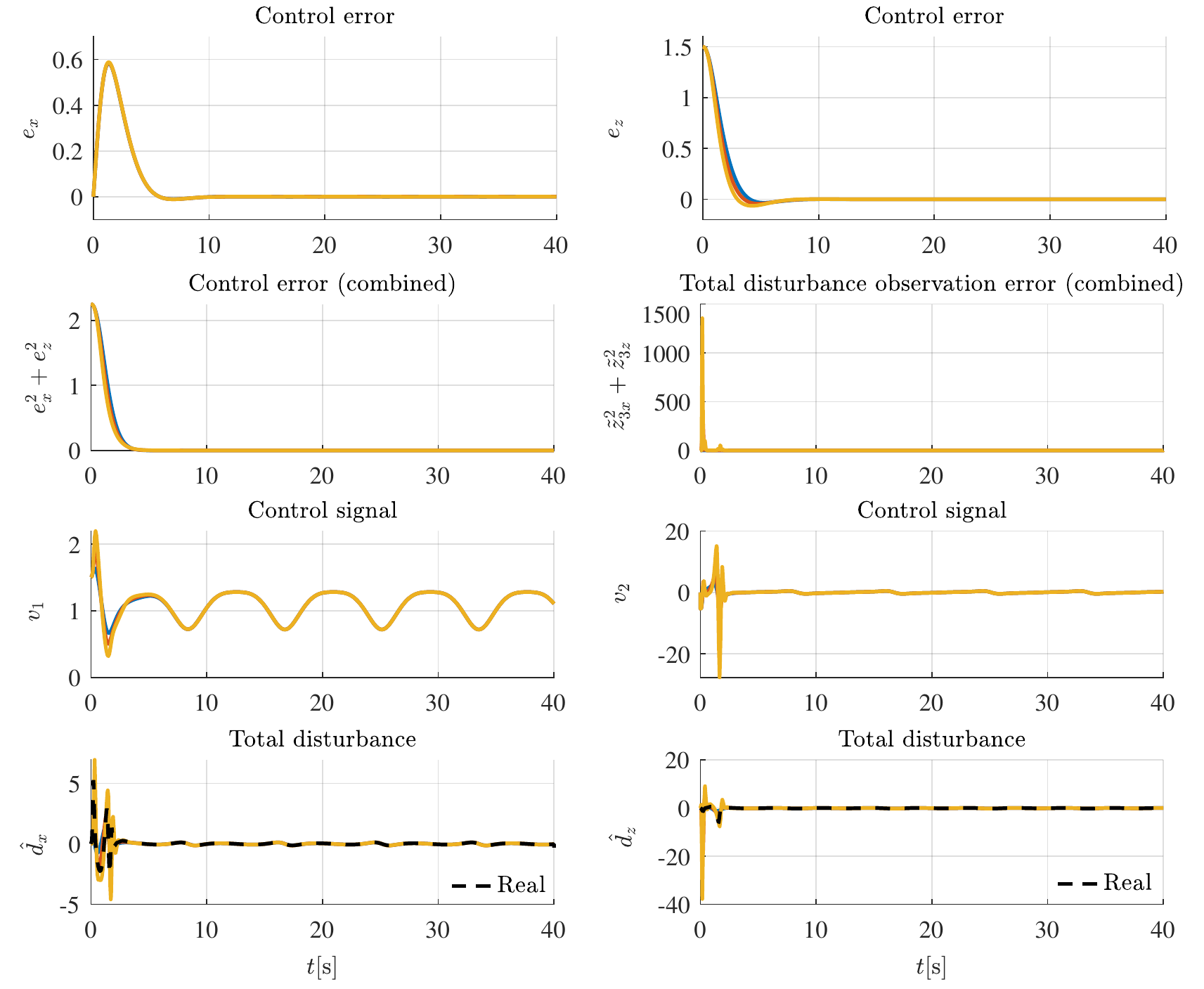}
    \caption{\textcolor{black}{Results of test \textit{Sim2A} (i.e. without measurement noise: $w(t)=0$).}}
    \label{fig:Sim2A}
\end{figure}

\begin{figure}[t]
  \centering
    \includegraphics[width=\textwidth]{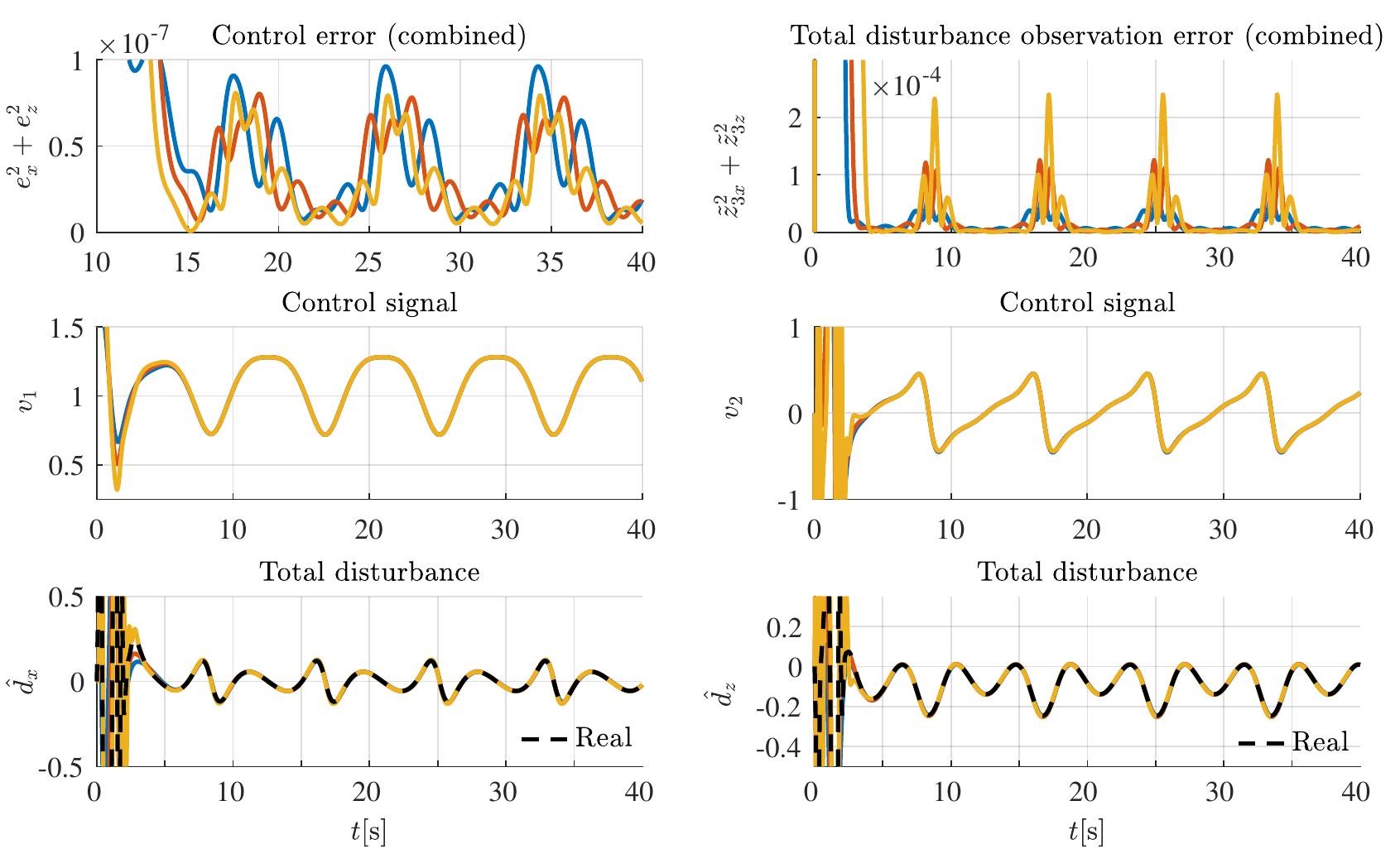}
    \caption{\textcolor{black}{Zoomed in results of test \textit{Sim2A} (cf.~Fig.~\ref{fig:Sim2A}).}}
    \label{fig:Sim2Azoom}
\end{figure}

The results of the noiseless case \textit{Sim2A} are gathered in Figs.~\ref{fig:Sim2A} and~\ref{fig:Sim2Azoom}. Even though a more complex system has been used here, comparing to the one from Sect.~\ref{ref:genericEx}, similar observations can be made. Despite the fact that the cascade ESO-based designs (i.e. for $p=2$ and $p=3$) have significantly lower observer bandwidths than standard approach, all of the tested observer structures provide visually comparable results of the total disturbance estimation error and energy consumption.

\begin{figure}[p]
  \centering
    \includegraphics[width=0.75\textwidth]{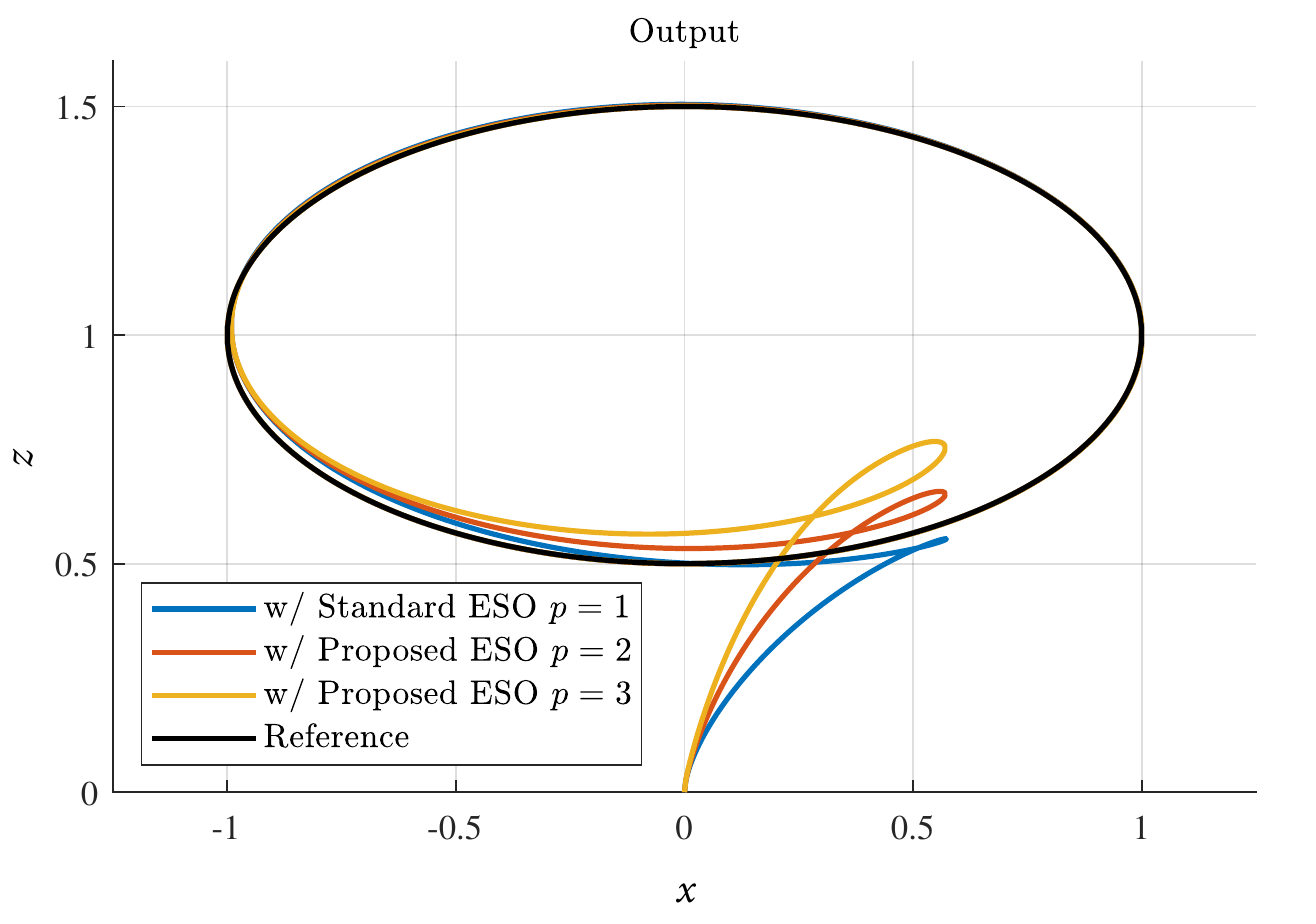}\\
    \includegraphics[width=\textwidth]{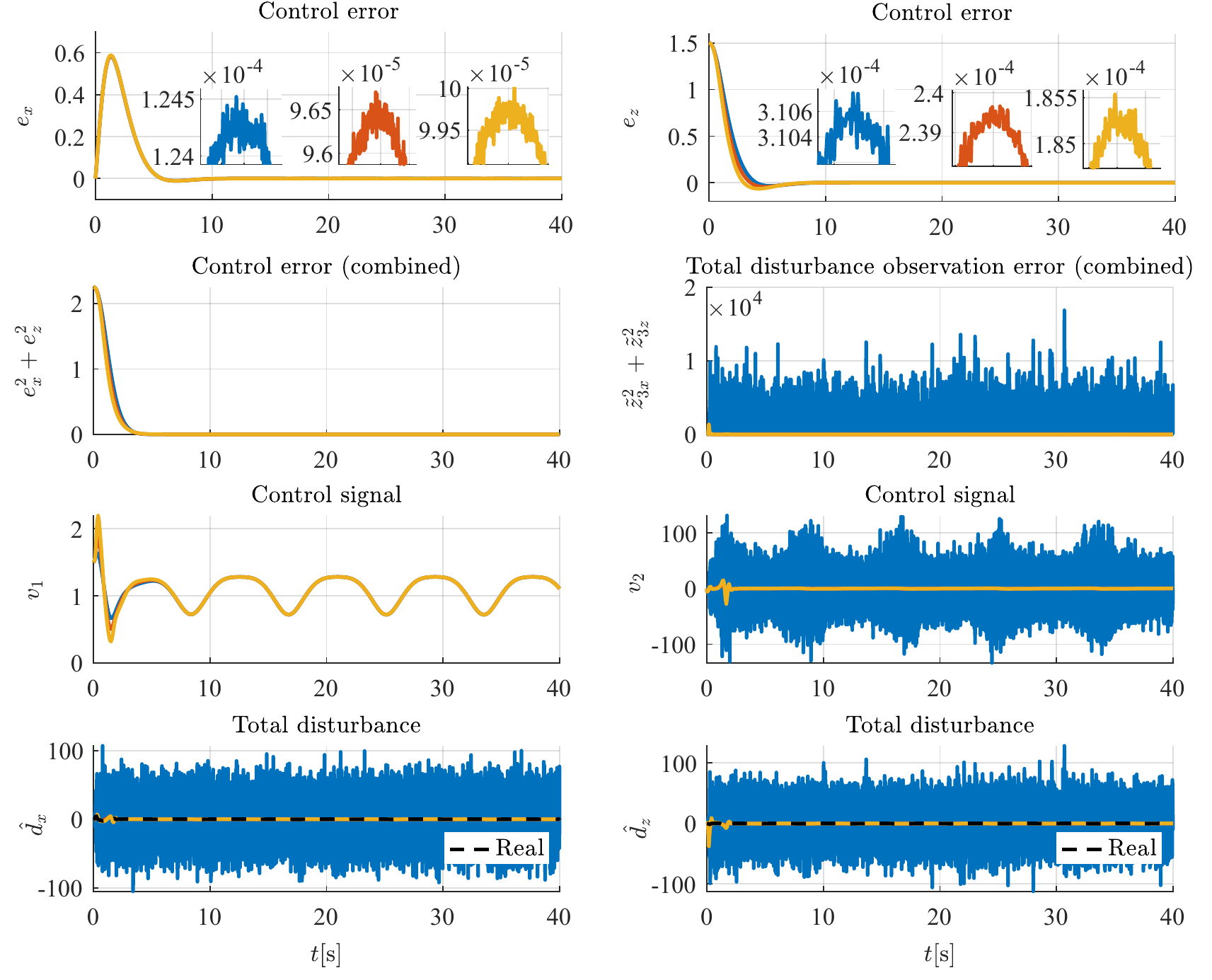}
    \caption{\textcolor{black}{Results of test \textit{Sim2B} (i.e. with measurement noise: $w(t)\neq 0$).}}
    \label{fig:Sim2B}
\end{figure}

\begin{figure}[t]
  \centering
    \includegraphics[width=\textwidth]{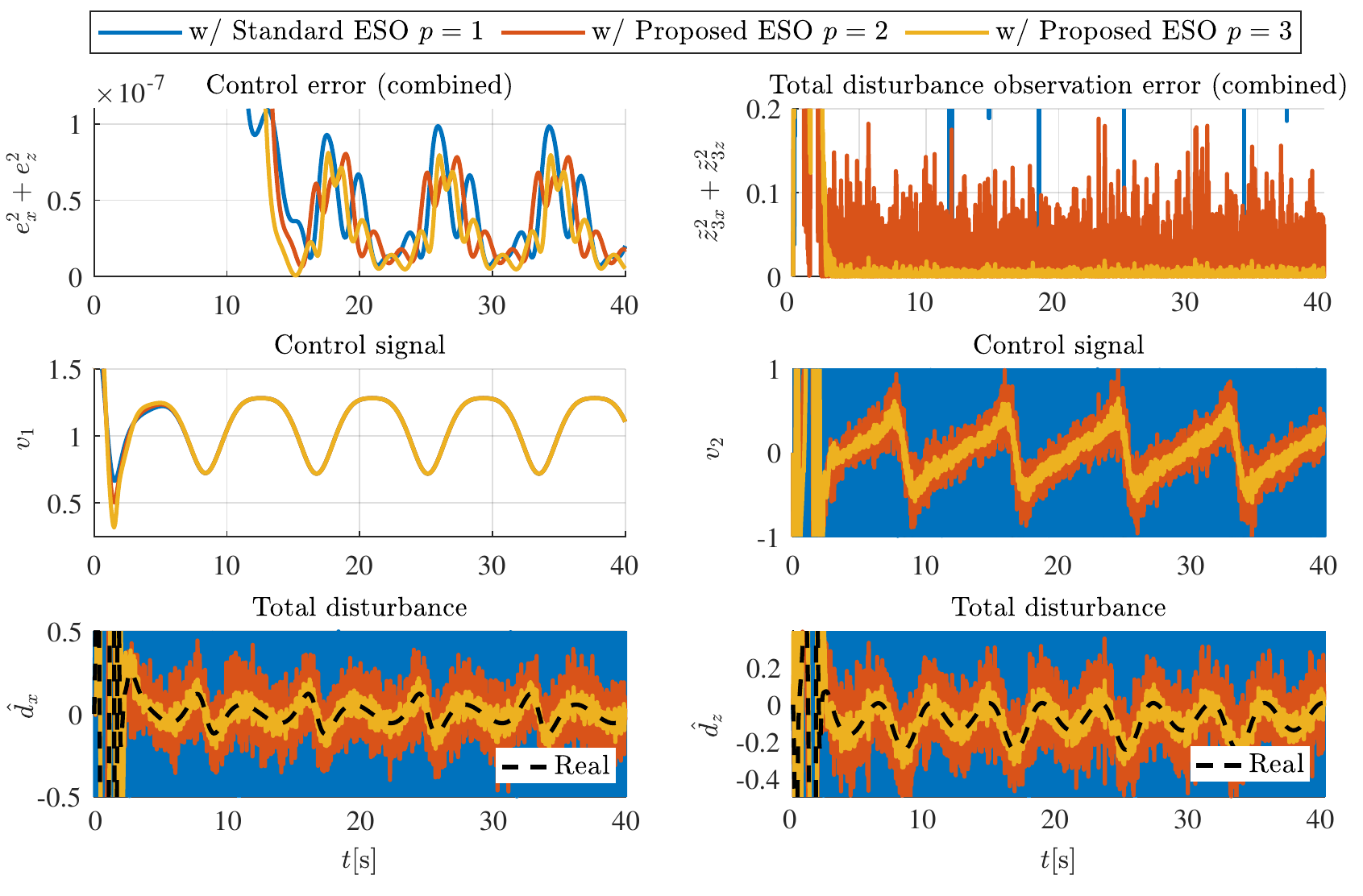}
    \caption{\textcolor{black}{Zoomed in results of test \textit{Sim2B} (cf.~Fig.~\ref{fig:Sim2B}).}}
    \label{fig:Sim2Bzoom}
\end{figure}

The outcomes of the noisy case \textit{Sim2B} are presented in Figs.~\ref{fig:Sim2B} and~\ref{fig:Sim2Bzoom}. \color{black}From the obtained results, one can notice that despite the presence of measurement noise visible on zooms embedded in the plot presenting the control errors $e_x$ and $e_z$ in Fig.~\ref{fig:Sim2B}\color{black}, the proposed cascade ESO provides improvement in the noise attenuation over standard ESO, especially in the quality of total disturbance reconstruction and, consequently, in the control signal $v_2$. With the extra layer of cascade ($p=3$), the amplitude of noise is further reduced. The measurement noise does not visually impact signal $v_1$, since it origins from the doubly integrated signals $u_x$ and $u_z$, as seen in~\eqref{eq:controlsPVTOL}.

\color{black}

\section{Conclusions}

From the obtained results, one can deduce that the compromise between speed/accuracy of state reconstruction and noise amplification is still present, however, with the cascade ESO one gains extra degree of freedom in designing high-bandwidth observers while minimizing the amplification of output noise. This reveals great practical potential of the proposed method in high-performance control of systems subjected to non-negligible measurement noise. The potential drawbacks of the proposed cascade ESO-based design (like extra tuning parameters, lack of a systematic method of cascade level selection and cascade ESO tuning) need to be further investigated.






\appendix
\color{black}
\section{Input-to-State (ISS) stability}

Following work \cite{mironchenko2016}, Definition 4.7 from \cite{khalil2002} and Assumption 12.3 from \cite{isidori2017}, a following system
\begin{align}
    \stateVectorDerivative(t) = f(\stateVector,\pmb{u}_1, \pmb{u}_2, t),
    \label{eq:isssystem}
\end{align}
defined on $\stateVector\in\mathcal{X}\subset\realNumbers^{n}$, $\pmb{u}_1\in\mathcal{U}_1\subset\realNumbers^{m_1}$, $\pmb{u}_2\in\mathcal{U}_2\subset\realNumbers^{m_2}$, and $t\in[0,\infty)$ is locally input-to-state stable (ISS) on some bounded sets $\mathcal{X}, \ \mathcal{U}_1, \ \mathcal{U}_2$ if there exists a function $\beta$ of class $\mathcal{KL}$, and functions $\gamma_1$, $\gamma_2$ of class $\mathcal{K}$, such that for any initial conditions $\stateVector(0)\in\mathcal{X}$ and any bounded inputs $\pmb{u}_1(t)$, $\pmb{u}_2(t)$, solution $\pmb{x}(t)$ exists for all $t$, satisfying
\begin{align}
    \|\stateVector(t)\| \leq \beta(\|\stateVector(0)\|,t)+\gamma_1\left(\sup_{t\geq0}\|\pmb{u}_1(t)\|\right)+\gamma_2\left(\sup_{t\geq0}\|\pmb{u}_2(t)\|\right),
\end{align}
and implying the asymptotic relation
\begin{align}
    \limsup_{t\rightarrow0}\|\stateVector(t)\| \leq \gamma_1\left(\sup_{t\geq0}\|\pmb{u}_1(t)\|\right)+\gamma_2\left(\sup_{t\geq0}\|\pmb{u}_2(t)\|\right).
    \label{eq:limsupRelation}
\end{align}

According to Theorem 4.19 from \cite{khalil2002}, if the continuously differentiable function $V(\stateVector):\mathcal{X}\rightarrow\realNumbers$ satisfies
\begin{align}
 \alpha_{1}(\|\stateVector\|)\leq V(\stateVector) \leq \alpha_2(\|\stateVector\|),
\end{align}
for some $\mathcal{K}_{\infty}$-class functions $\alpha_1$ and $\alpha_2$, and
\begin{align}
    \dot{V}(\stateVector,\pmb{u}_1,\pmb{u}_2,t) \leq -\alpha_3(\stateVector) \quad \textrm{for} \quad \|\stateVector\|\geq\phi_1(\|\pmb{u}_1\|)+\phi_2(\|\pmb{u}_2\|)>0,
\end{align}
for a continuous positive-definite function $\alpha_3(\stateVector):\mathcal{X}\rightarrow\positiveRealNumbers$, and a class $\mathcal{K}$ functions $\phi_1, \ \phi_2$, then, system \eqref{eq:isssystem} is locally ISS with respect to the inputs $\pmb{u}_1$ and $\pmb{u}_2$, satisfying \eqref{eq:limsupRelation} for functions $\gamma_1(\cdot) = \alpha_1^{-1}(\alpha_2(\phi_1(\cdot)))$, and $\gamma_2(\cdot) = \alpha_1^{-1}(\alpha_2(\phi_2(\cdot)))$.

\color{black}


\bibliography{mybibfile}
\end{document}

%% file: definitions.tex
\newcommand{\zeroMatrix}{\pmb{0}}
\newcommand{\identityMatrix}{\pmb{I}}
\newcommand{\realNumbers}{\mathbb{R}}
\newcommand{\positiveRealNumbers}{\mathbb{R}_+}
\newcommand{\integerNumbers}{\mathbb{Z}}
\newcommand{\module}[1]{\left\|{#1}\right\|}
\newcommand{\abs}[1]{|{#1}|}

\newcommand{\minEigenvalue}[1]{\lambda_{\textrm{min}}({#1})}
\newcommand{\maxEigenvalue}[1]{\lambda_{\textrm{max}}({#1})}

\newcommand{\stateVector}{\pmb{x}}

\newcommand{\stateVectorElement}[1]{{x}_{#1}}
\newcommand{\stateVectorDerivative}{\dot{\pmb{x}}}
\newcommand{\systemOutput}{y}
\newcommand{\controlSignal}{u}
\newcommand{\virtualControlSignal}[1]{v}

\newcommand{\virtualControlSignalIteration}[1]{v}
\newcommand{\controlSignalField}{g}
\newcommand{\systemNonlinearDynamics}{f}
\newcommand{\controlSignalFieldEstimate}{\hat{g}}
\newcommand{\disturbance}{d^*}
\newcommand{\disturbanceDerivative}{\dot{d}^*}

\newcommand{\differentiableFunctionSet}[1]{\mathcal{C}^{{#1}}}
\newcommand{\totalDisturbance}{d}
\newcommand{\totalDisturbanceDerivative}{\dot{d}}
\newcommand{\totalDisturbanceEstimate}{\hat{d}}
\newcommand{\totalDisturbanceObservationError}{\tilde{d}}
\newcommand{\totalDisturbanceObservationErrorEstimate}{\hat{\tilde{d}}}

\newcommand{\observerStageState}[1]{\pmb{\xi}_{#1}}
\newcommand{\observerStageStateElement}[1]{{\xi}_{#1}}
\newcommand{\observerStageStateElementEstimate}[1]{\hat{\xi}_{#1}}
\newcommand{\observerStageStateDerivative}[1]{\dot{\pmb{\xi}}_{#1}}

\newcommand{\extendedState}{\pmb{z}}
\newcommand{\extendedStateDerivative}{\dot{\pmb{z}}}
\newcommand{\extendedStateEstimate}{\hat{\pmb{z}}}

\newcommand{\stateMatrix}[1]{\pmb{A}_{#1}}

\newcommand{\outputVector}[1]{\pmb{c}_{#1}}
\newcommand{\inputVector}[1]{\pmb{b}_{#1}}
\newcommand{\inputVectorExtendedState}[1]{\pmb{d}_{#1}}

\newcommand{\measurementNoise}{w}
\newcommand{\observerGainVector}[1]{\pmb{l}_{#1}}

\newcommand{\observerGainVectorElementCoefficient}[1]{{\kappa}_{{#1}}}

\newcommand{\observerStageBandwidth}[1]{\omega_{o #1}}

\newcommand{\extendedStateObservationError}{\tilde{\pmb{z}}}

\newcommand{\observationErrorStateMatrix}[1]{\pmb{H}_{#1}}

\newcommand{\observerBandwidthMultiplier}{\alpha}
\newcommand{\stateTransformationMatrix}[1]{\pmb{\Lambda}_{#1}}
\newcommand{\transformedStateStateMatrix}[1]{\pmb{H}^*}
\newcommand{\transformedState}[1]{\pmb{\zeta}_{#1}}
\newcommand{\transformedStateDerivative}[1]{\dot{\pmb{\zeta}}_{#1}}
\newcommand{\observerStageObservationError}[1]{\tilde{\pmb{\xi}}_{#1}}
\newcommand{\observerStageObservationErrorDerivative}[1]{\dot{\tilde{\pmb{\xi}}}_{#1}}
\newcommand{\observerStageLyapunovFunction}[1]{V_{#1}}
\newcommand{\observerStageLyapunovFunctionDerivative}[1]{\dot{V}_{#1}}
\newcommand{\observerStageMajorizationConstant}[1]{\nu_{#1}}
\newcommand{\maximalGainCoefficient}{\kappa_{\textrm{max}}}
\newcommand{\residualObservationErrorMatrix}[1]{\pmb{\Gamma}_{#1}}
\newcommand{\measurementNoiseVector}[1]{\pmb{\delta}_{#1}}

\newcommand{\lyapunovEquationSolutionMatrix}[1]{\pmb{P}}

\newcommand{\kappaFunction}{\mathcal{K}}

\newcommand{\systemStateDomain}{\mathcal{D}_x}

\newcommand{\totalDisturbanceDerivativeDomain}{\mathcal{D}_{\totalDisturbanceDerivative}}
\newcommand{\externalDisturbanceDerivativeDomain}{\mathcal{D}_{\dot{d}^*}}
\newcommand{\externalDisturbanceDomain}{\mathcal{D}_{{d}^*}}

\newcommand{\measurementNoiseDomain}{\mathcal{D}_\measurementNoise}
\newcommand{\totalDisturbanceDerivativeUpperBound}{r_{\totalDisturbanceDerivative}}
\newcommand{\measurementNoiseUpperBound}{r_{\measurementNoise}}

\newcommand{\issFactor}{\gamma}

\newcommand{\pvtolX}{x}
\newcommand{\pvtolXDerivative}{\dot{x}}
\newcommand{\pvtolXSecondDerivative}{\ddot{x}}
\newcommand{\pvtolXThirdDerivative}{{x}^{(3)}}
\newcommand{\pvtolXFourthDerivative}{{x}^{(4)}}
\newcommand{\pvtolY}{z}
\newcommand{\pvtolYDerivative}{\dot{z}}
\newcommand{\pvtolYSecondDerivative}{\ddot{z}}
\newcommand{\pvtolYThirdDerivative}{{z}^{(3)}}
\newcommand{\pvtolYFourthDerivative}{{z}^{(4)}}
\newcommand{\pvtolOrientation}{\theta}
\newcommand{\pvtolOrientationDerivative}{\dot{\theta}}
\newcommand{\pvtolOrientationSecondDerivative}{\ddot{\theta}}
\newcommand{\pvtolThrust}{v_1}
\newcommand{\pvtolThrustDerivative}{\dot{v}_1}
\newcommand{\pvtolThrustSecondDerivative}{\ddot{v}_1}
\newcommand{\pvtolMomentum}{v_2}
\newcommand{\pvtolWindThrust}{d_1^*}
\newcommand{\pvtolWindMomentum}{d_2^*}
\newcommand{\pvtolWindThrustDerivative}{\dot{d}_1^*}

\newcommand{\pvtolWindThrustSecondDerivative}{\ddot{d}_1^*}
\newcommand{\pvtolWindMomentumSecondDerivative}{\ddot{d}_2^*}

\newcommand{\pvtolMatrix}{M}
\newcommand{\pvtolVirtualControlSignalX}{u_x}
\newcommand{\pvtolVirtualControlSignalY}{u_z}

\newcommand{\pvtolTotalDisturbanceX}{d_x}
\newcommand{\pvtolTotalDisturbanceY}{d_z}

\renewcommand{\arraystretch}{1.2}